\documentclass[%
 reprint,
superscriptaddress,
nofootinbib,
 amsmath,amssymb,
 aps,
]{revtex4-2}
\usepackage{graphicx}
\usepackage{dcolumn}
\usepackage{bm}
\usepackage{mathtools}
\usepackage{color}
\usepackage[colorlinks=true,linkcolor=blue,citecolor=blue]{hyperref}

\newcommand{\Etmx}{\ensuremath{E_{3\rm max}}}
\newcommand{\emax}{\ensuremath{e_{\rm max}}}
\newcommand{\com}{\ensuremath{\rm cm}}
\newcommand{\rel}{\ensuremath{\rm rel}}
\newcommand{\fm}{{\rm fm}}
\newcommand{\sixj}[6]{\left\{
    \begin{array}{ccc}
      #1 & #2 & #3 \\
      #4 & #5 & #6
    \end{array}
\right\}}

\begin{document}

\preprint{APS/123-QED}

\title{Converged ab initio calculations of heavy nuclei}

\author{T.~Miyagi}
\email{tmiyagi@triumf.ca}
\affiliation{TRIUMF, 4004 Wesbrook Mall, Vancouver, BC V6T 2A3, Canada}
\author{S.~R.~Stroberg}%
\email{stroberg@uw.edu}
\affiliation{Department of Physics, University of Washington, Seattle, Washington 98195, USA}%
\author{P.~Navr\'atil}
\email{navratil@triumf.ca}
\affiliation{TRIUMF, 4004 Wesbrook Mall, Vancouver, BC V6T 2A3, Canada}
\author{K.~Hebeler}
\email{kai.hebeler@physik.tu-darmstadt.de}
\affiliation{Technische Universit\"at Darmstadt, 64289 Darmstadt, Germany}
\affiliation{ExtreMe Matter Institute EMMI, GSI Helmholtzzentrum f\"ur Schwerionenforschung GmbH, 64291 Darmstadt, Germany}%
\affiliation{Max-Planck-Institut f\"ur Kernphysik, Saupfercheckweg 1, 69117 Heidelberg, Germany}
\author{J.~D.~Holt}
\email{jholt@triumf.ca}
\affiliation{TRIUMF, 4004 Wesbrook Mall, Vancouver, BC V6T 2A3, Canada}
\affiliation{Department of Physics, McGill University, 3600 Rue University, Montr\'eal, QC H3A 2T8, Canada}%

\begin{abstract}
We propose a novel storage scheme for three-nucleon (3N) interaction matrix elements relevant for the normal-ordered two-body approximation used extensively in ab initio calculations of atomic nuclei.
This scheme reduces the required memory by approximately two orders of magnitude, which allows the generation of 3N interaction matrix elements with the standard truncation of $E_{\rm 3max}=28$, well beyond the previous limit of 18. We demonstrate that this is sufficient to obtain the ground-state energy of $^{132}$Sn converged to within a few MeV with respect to the $E_{\rm 3max}$ truncation.
In addition, we study the asymptotic convergence behavior and perform extrapolations to the un-truncated limit.
Finally, we investigate the impact of truncations made when evolving free-space 3N interactions with the similarity renormalization group.
We find that the contribution of blocks with angular momentum $J_{\rel}>9/2$ to the ground-state energy is dominated by a basis-truncation artifact which vanishes in the large-space limit, so these computationally expensive components can be neglected.
For the two sets of nuclear interactions employed in this work, the resulting binding energy of $^{132}$Sn agrees with the experimental value within theoretical uncertainties.
This work enables converged ab initio calculations of heavy nuclei.
\end{abstract}

\maketitle


\section{\label{sec:intro}Introduction}
With recent progress in constructing two- (NN) and three-nucleon (3N) interactions~\cite{Epelbaum2009,Machleidt2011}, solving the nuclear many-body problem~\cite{Navratil2016,Carlson2015,Lee2009,Hagen2014,Dickhoff2004,Hergert2016,Tichai2020}, and rapid increases in computational power, the range of applicability of ab initio calculations of atomic nuclei has exploded over the past decade~\cite{Hergert2020}.
On the side of nuclear interactions, it has become clear that a consistent treatment of NN scattering and finite nuclei requires the inclusion of 3N forces~\cite{Pieper2002,Navratil2003,Otsuka2010,Hebeler2015,Hebeler2020}, where chiral effective field theory~\cite{Epelbaum2009,Machleidt2011,Hammer2020} provides a path to a consistent and systematic treatment.

On the many-body side, polynomially scaling methods, such as coupled-cluster theory~\cite{Hagen2014}, self-consistent Green's functions~\cite{Dickhoff2004}, and in-medium similarity renormalization group (IMSRG)~\cite{Hergert2016} have been used to treat systems of up to $A\sim 100$ particles~\cite{Morris2018,Arthuis2020,Gysbers2019}.
In all of these calculations, the wave function is expanded on a set of basis functions---typically the eigenstates of the harmonic oscillator---and the NN and 3N matrix elements in that basis are needed as an input.
The number of single-particle basis states in a calculation is given by the truncation $e=2n+\ell \leq e_{\max}$, with the radial quantum number $n$ and angular momentum $l$.
Achieving convergence in both the infrared (IR) and ultraviolet (UV) for medium-mass nuclei typically requires $\emax\gtrsim 12$.
At even $e_{\max}=12$, however, storing the full set of 3N matrix elements would require approximately 10 TB of memory with single-precision floating point numbers, which considerably exceeds the available RAM per node on a typical supercomputer.
It is therefore necessary to impose some additional truncation on the 3N matrix elements, typically taken as $e_1+e_2+e_3\leq \Etmx$.
Ideally, the value of $\Etmx$ is increased until convergence is achieved for a given observable.

The current limit of $\Etmx\lesssim 18$ is the primary bottleneck preventing ab initio calculations from reaching much beyond $A \sim 100$~\cite{Binder2014,Lascar2017,Arthuis2020,Mane20Cd}.
Overcoming this limit would significantly increase the reach of ab initio theory, e.g. to searches for physics beyond the standard model using heavy isotopes of xenon, tellurium, cesium, or mercury~\cite{Adams2020,Anton2019,Aprile2019,Suzuki2019,Akerib2016,Gando2016,Engel2013}.
Furthermore, potential controlled calculations of $^{208}$Pb would provide the best experimentally accessible link between finite nuclei and nuclear matter, particularly in light of recently reported parity-violating electron scattering experiments~\cite{Horowitz2001,Roca-Maza2011,Abrahamyan2012}.
Ab initio predictions would even be possible for the astrophysically relevant, but experimentally challenging, $N$=126 region below $^{208}$Pb~\cite{Grawe2007,Watanabe2015,Savard2020}.

One way to overcome this limitation is to apply an importance truncation and/or tensor factorization~\cite{Tichai2019,Tichai2019a} to the 3N matrix elements, which would dramatically reduce the required RAM while retaining sufficient accuracy.
Before resorting to these techniques, however, we observe that the most of today's practical calculations are based on the normal-ordered two-body (NO2B) approximation~\cite{Roth2012}.
This means we do not need the full set of 3N matrix elements in actual applications, particularly in the heavy-mass region. In this work, we demonstrate the efficiency of generating and storing only those combinations of 3N matrix elements involved in the NO2B approximation and discuss the $\Etmx$ convergence of heavy nuclei around $^{132}$Sn.

The structure of this paper is as follows. In Sec.~\ref{sec:no2b}, we introduce a novel procedure to store the 3N matrix elements relevant to the NO2B approximation.
In Sec.~\ref{sec:conv}, the asymptotic behavior with respect to $\Etmx$ is discussed.
In Sec.~\ref{sec:results}, we demonstrate large $\Etmx$ calculations around $^{132}$Sn, using the well-established NN+3N 1.8/2.0 (EM) interaction~\cite{Hebeler2011}.
We also discuss the uncertainty from free-space 3N similarity renormalization group (SRG) evolution and present results for $^{132}$Sn with the chiral NN+3N(lnl) interaction~\cite{Soma2019}.
Finally, we conclude in Sec.~\ref{sec:conclusion}.

\section{\label{sec:no2b}Calculation of 3N matrix elements}
\begin{figure}[t]
    \centering
    \includegraphics[clip,width=\columnwidth]{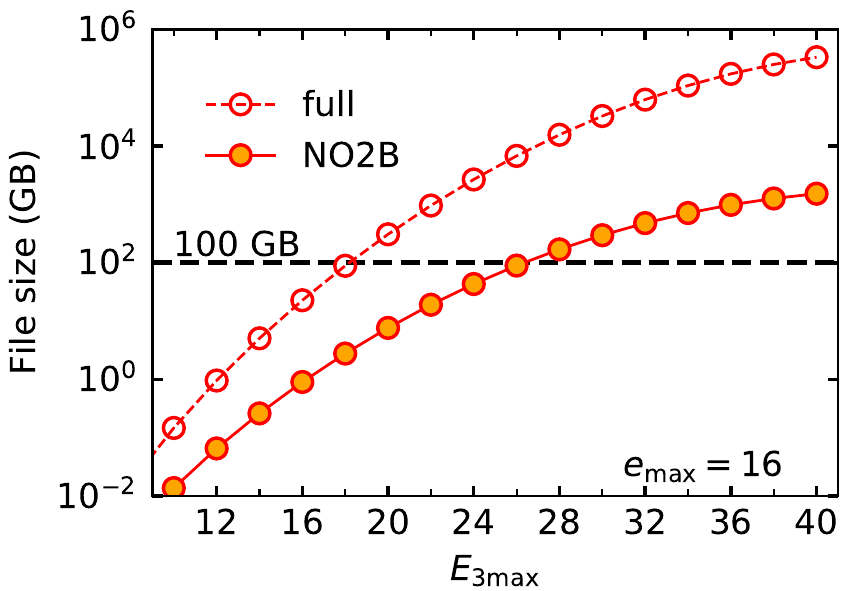}
    \caption{File size of the three-body matrix elements with the single-precision floating point numbers. The horizontal dashed line indicates 100 GB, which is a typical limit of the memory per node in usual work stations.}
    \label{fig:memory}
\end{figure}

In Figure~\ref{fig:memory} we show the estimated file size of the 3N matrix elements as a function of $E_{\rm 3max}$ for a fixed $\emax=16$.
The curve ``full'' illustrates that the typical basis-size limit is approximately $E_{3\max}=16-18$ for a memory limit of about 100 GB. This limit, however, is typically not sufficient to obtain converged results for nuclei beyond $A=100$ as discussed in Refs.~\cite{Binder2014,Lascar2017,Arthuis2020,Mane20Cd,Whit20tin}, and which we also demonstrate below.
Towards heavier systems, the contributions of the residual 3N interactions is expected to be comparable to the truncation error of the many-body method~\cite{Binder2013}.
Since the memory requirement for storing the full set of 3N matrix elements is prohibitive, we instead aim to exploit the simplifications offered by the NO approximation.
In order to identify the minimal subset of 3N matrix elements for the NO2B Hamiltonian, we begin by reviewing the normal-ordering procedure.

\subsection{NO2B 3N matrix elements}
Our starting Hamiltonian in second-quantized form is
\begin{align}
H = \sum_{p'p}& t_{p'p} a^{\dag}_{p'}a_{p}
 + \frac{1}{4} \sum_{pp'qq'} V^{\rm NN}_{p'q'pq} a^{\dag}_{p'}a^{\dag}_{q'}a_{q}a_{p} \notag \\
 &+ \frac{1}{36} \sum_{pp'qq'rr'} V^{\rm 3N}_{p'q'r'pqr} a^{\dag}_{p'}a^{\dag}_{q'}a^{\dag}_{r'}a_{r}a_{q}a_{p},
\end{align}
where $t_{p'p}$, $V^{\rm NN}_{p'q'pq}$, and $V^{\rm 3N}_{p'q'r'pqr}$ are the one-, two-, and three-body matrix elements, respectively.
The index $p$ labels the single-particle orbit with quantum numbers $\{n_{p}, \ell_{p}, j_{p}, m_{p}, t_{z_{p}}\}$ corresponding to the radial quantum number, orbital angular momentum, total angular momentum, total angular momentum projection, and isospin projection, respectively.
Performing normal ordering with respect to a reference state characterized by a one-body density matrix $\rho_{p'p}=\langle a^{\dagger}_{p'}a_{p}\rangle$
and discarding the residual 3N part, we obtain the NO2B Hamiltonian:
\begin{equation}
\begin{aligned}
H^{\rm (NO2B)}=
E_{0} &+ \sum_{p'p} f_{p'p} \{ a^{\dag}_{p'}a_{p}\} \\
+& \frac{1}{4} \sum_{pp'qq'} \Gamma_{p'q'pq}
\{ a^{\dag}_{p'}a^{\dag}_{q'}a_{q} a_{p}\}\, ,
\end{aligned}
\end{equation}
where the braces $\{\ldots\}$ indicate that the enclosed string of creation and annihilation operators are normal-ordered with respect to the used reference state. The Hamiltonian is now expressed in terms of a zero-body part
\begin{equation}
\begin{aligned}
E_{0} = \sum_{p'p} \rho_{p'p} t_{p'p} &+ \frac{1}{2} \sum_{pp'qq'}
\rho_{p'p}\rho_{q'q} V^{\rm NN}_{p'q'pq}\\
+\frac{1}{6} &\sum_{pp'qq'rr'}
\rho_{p'p}\rho_{q'q} \rho_{r'r} V^{\rm 3N}_{p'q'r'pqr},
\end{aligned}
\end{equation}
a normal-ordered one-body part
\begin{equation}
f_{p'p} = t_{p'p} + \sum_{q'q} \rho_{q'q} V^{\rm NN}_{p'q'pq} + \frac{1}{2} \sum_{qq'rr'} \rho_{q'q} \rho_{r'r} V^{\rm 3N}_{q'r'p'qrp},
\end{equation}
and a normal-ordered two-body part
\begin{equation}\label{eq:GammaM}
\Gamma_{p'q'pq} = V^{\rm NN}_{p'q'pq} + \sum_{r'r} \rho_{r'r} V^{\rm 3N}_{p'q'r'pqr}.
\end{equation}
The accuracy of the NO2B approximation has been investigated for ground state energies~\cite{Roth2012,Binder2013,Djarv2021}, where it was found that by $^{16}$O the error is at the level of 1\% of the binding energy.
With increasing mass number, this error should decrease as a fraction of the total binding energy\footnote{
The approximation also breaks translational invariance~\cite{Djarv2021}, but this is only important for light nuclei (i.e. $A\lesssim16$), where the NO2B truncation is not necessary and convergence in $\Etmx$ can be obtained by conventional methods.}.

If the one-body density matrix $\rho_{pp'}$ is rotationally invariant and conserves parity and isospin projection, it must satisfy
$(\ell_{p'},j_{p'},m_{p'},t_{z_{p'}}) = (\ell_{p},j_{p},m_{p},t_{z_{p}})$,
and the number of required 3N matrix elements is drastically smaller than that of the original full set.
This condition is satisfied for single-reference calculations (e.g. coupled cluster, self-consistent Green's function, IMSRG, HF-MBPT) with a closed-shell reference, as well as for particle-attached and particle-removed methods~\cite{Hagen2014}, and the ensemble normal ordering reference used in the valence-space IMSRG~\cite{Stroberg2017}.
On the other hand, for broken-symmetry~\cite{Signoracci2015,Novario2020} or  multi-configurational~\cite{Hergert2013,Gebrerufael2016} references necessary to describe e.g. well-deformed nuclei, this would constitute an additional approximation.

Furthermore, in the practically used $JT$-coupled representation, we can sum up the 3N total angular momentum dependence.
This can be seen in the $J$-coupled expression for the normal-ordered matrix element, the full expressions for which are provided in Appendix~\ref{appendix:NO}.
Here we show only the contributions from the three-nucleon interactions $V^{\rm 3N}$ (with the notation $[x]\equiv 2x+1$, and using un-normalized matrix elements):
\begin{subequations}\label{eq:NO2BJ3b}
\begin{align}
    &E_0~[{\rm 3N}] =  \frac{1}{6} \sum_{\substack{p'q'r'\\pqr}} \rho_{p'p}\rho_{q'q}\rho_{r'r} \sum_{J_{pq}J}[J] \; V^{J_{pq}J_{pq}J}_{p'q'r'pqr} \\
    &f_{p'p}~[{\rm 3N}] =
     \frac{1}{2}\sum_{qq'rr'} \rho_{q'q}\rho_{r'r} \sum_{J_{qr}J}\frac{[J]}{[j_{p}]} V^{J_{qr}J_{qr}J}_{q'r'p'qrp} \\
    &\Gamma^{J_{pq}}_{p'q'pq}~[{\rm 3N}] =  \sum_{rr'} \rho_{r'r} \sum_{J}\frac{[J]}{[J_{pq}]} V^{J_{pq}J_{pq}J}_{p'q'r'pqr}\, .
\end{align}
\end{subequations}
We can see that in Eq.~\eqref{eq:NO2BJ3b} all terms depend on $V^{\rm 3\textbf{}N}$ through the quantity
\begin{equation}   \label{eq:elem_no2b}
    \mathcal{V}_{p'q'r'pqr}^{J_{pq}}
    \equiv
    \sum_{J}[J] V_{p'q'r'pqr}^{J_{pq}J_{pq}J}
    \tilde{\delta}_{r'r} \, ,
\end{equation}
where the symbol $\tilde{\delta}_{r'r}$ is shorthand for all the quantum numbers which are conserved by the one-body density matrix $\rho_{r'r}$.
If, instead of the full $V^{\rm 3N}$, we only store the quantity (\ref{eq:elem_no2b}), we obtain the curve ``NO2B" in Fig.~\ref{fig:memory}, allowing us to access $\Etmx=26$ $(28)$ using single- (half-) precision floating point numbers.

Note that for a Hartree-Fock (HF) calculation, we only need the combination
\begin{equation}\label{eq:Vmonopole}
    \overline{\mathcal{V}}_{p'q'r'pqr} \equiv \sum_{J_{pq}}
    \mathcal{V}_{p'q'r'pqr}^{J_{pq}}
    \tilde{\delta}_{p'p}
    \tilde{\delta}_{q'q}.
\end{equation}
The number of matrix elements \eqref{eq:Vmonopole} is sufficiently low that we can store the full $\overline{\mathcal{V}}_{p'q'r'pqr}$ without any $\Etmx$ truncation.
However, we find that the HF part of the calculation converges at lower $\Etmx$ than the beyond-mean-field corrections, which is why we store the q\textbf{}uantity $\mathcal{V}^{J_{pq}}_{p'q'r'pqr}$.

A similar idea was employed in Ref.~\cite{Binder2014}.
However, in that work an iterative procedure was adopted in which a HF calculation was performed at a manageable $\Etmx=14$, and then the lab-frame matrix elements necessary for the NO2B approximation for larger $\Etmx$ were computed from the relative-basis matrix elements, and the procedure was iterated until self consistency was attained.
In our approach, the transformation to the lab frame is performed once, and the resulting matrix elements $\mathcal{V}$ are written to disk and can be used for future calculations of any desired nucleus, without the need for iteration.

\subsection{Transformation to single-particle coordinate}
Although we can compress the file size by calculating only the NO2B relevant matrix elements via Eq.~(\ref{eq:elem_no2b}), we still need an efficient way to perform the transformation from the three-body Jacobi basis to single-particle basis.
Originally, this transformation was derived for the three-nucleon single-particle $m$-scheme basis~\cite{Navratil2003,Nogga2006}. Memory requirements for the $m$-scheme storage limited calculations to $\Etmx{=}9$.
Later, a $j$-coupled storage scheme was introduced~\cite{Roth2011,Roth2014} that allowed calculations with $\Etmx\lesssim 18$ as discussed in the introduction with file size requirements illustrated in Fig.~\ref{fig:memory}.
Here, we specify the angular momentum coupling in detail. The antisymmetrized three-body states in the lab frame are defined as
\begin{equation}\label{eq:abcJ}
\begin{aligned}
    |pqr:J_{pq}T_{pq}JT& \rangle = \sqrt{6}\mathcal{A}
    \sum_{\{t_{z}\}} \mathcal{C}_{t_{z{p}}t_{z_{q}}T_{z{pq}}}^{t_{p} t_{q} T_{pq}}
    \mathcal{C}_{T_{z_{pq}}t_{z_{r}}T_z}^{T_{pq}t_r T} \\
    &\times\sum_{\{m\}}
    \mathcal{C}_{m_{p} m_{q} M_{pq}}^{j_{p} j_{q} J_{pq}}
    \mathcal{C}_{M_{pq}m_{r}M}^{J_{pq}j_{r}J}|p\rangle |q\rangle|r\rangle
    \end{aligned}
\end{equation}
with the antisymmetrizer
\begin{equation}
\mathcal{A} = \frac{1}{3!} ( 1 + P_{13}P_{12} + P_{12}P_{23} - P_{12} - P_{13} - P_{23}),
\end{equation}
defined in terms of the permutation operator $P_{ij}$.
In Eq.\eqref{eq:abcJ}, the symbol $\mathcal{C}$ indicate a Clebsch-Gordan coefficient.
A state in the antisymmetrized Jacobi basis is denoted
$|Ni J_{\rm rel}\rangle$, with total oscillator quanta $N$, total angular momentum $J_{\rm rel}$, and an additional quantum number $i$ to distinguish the states.
The transformation from the Jacobi basis to the lab frame may be expressed as

\begin{equation}\label{eq:3Ntrans}
\begin{aligned}
\langle &p'q'r': J_{p'q'}T_{p'q'} JT | V^{\rm 3N} | pqr:J_{pq}T_{pq}JT \rangle =\\ &6 \smashoperator{\sum_{\substack{NiN'i'\\N_{\com}L_{\com}J_{\rel}}}} 
\langle p'q'r': J_{p'q'} T_{p'q'} JT | N_{\com}L_{\com}N' i'J_{\rel} :JT \rangle
 \\ &\hspace{2em} \times
\langle N'i' J_{\rel}| V^{\rm 3N} | Ni J_{\rel} \rangle
\\ &\hspace{2em} \times
\langle N_{\com}L_{\com} NiJ_{\rel} :JT | pqr:J_{pq}T_{pq}JT\rangle.
\end{aligned}
\end{equation}
The quantity $\langle N_{\com}L_{\com} NiJ_{\rel} :JT | pqr:J_{pq}T_{pq}JT\rangle$ denotes the transformation coefficient. The quantum numbers $N_{\com}$ and $L_{\com}$ are the radial nodes and orbital angular momentum of the center-of-mass (c.m.)~motion.
The summations over $N,i,N',i'$ can be performed efficiently by matrix-matrix multiplication,
and the remaining summations over $N_{\rm cm}$, $L_{\rm cm}$ and $J_{\rm rel}$ can be computed manually.

The transformation coefficient can be calculated through the non-antisymmetrized Jacobi state:
\begin{equation}
\begin{aligned}
 \langle &N_{\com} L_{\com} NiJ_{\rel} :JT | pqr:J_{pq}T_{pq}JT\rangle
=\\
&\hspace{1em} \sum_{\alpha}
\langle N iJ_{\rel} | N \alpha J_{\rel}\rangle
\\ & \hspace{3em} \times
\langle N_{\com} L_{\com} N \alpha J_{\rel} :JT | pqr:J_{pq}T_{pq}JT\rangle.
\end{aligned}
\end{equation}
The index $\alpha$ labels the set of Jacobi quantum numbers
$\alpha = \{n_{12},l_{12},s_{12},j_{12},t_{12},n_{3},l_{3},j_{3} \}$.
The quantum numbers $\{n_{12}, l_{12}, s_{12}, j_{12}, t_{12}\}$ are used for the relative motion of nucleons 1 and 2, i.e., the nodal, orbital angular momentum, spin, total angular momentum, and total isospin quantum numbers, respectively.
Similarly, the quantum numbers $\{n_{3}, l_{3}, j_{3}\}$ correspond to the motion of nucleon 3 with respect to the c.m.~of nucleons 1 and 2.
Since the antisymmetrized state $|NiJ_{\rm rel} \rangle$ is an eigenstate of the antisymmetrizer $\mathcal{A}$, the coefficient $\langle N iJ_{\rm rel} | N \alpha J_{\rm rel}\rangle$ is also known as the coefficient of fractional parentage~\cite{Navratil1999,Navratil2000}.
The coefficient $\langle N_{\com} L_{\com} N \alpha J_{\rel} :JT | pqr:J_{pq}T_{pq}JT\rangle$ is known as the $T$-coefficient~\cite{Nogga2006,Roth2014}, and is the bottleneck of the calculation.
It turns out one can sum up three of the angular momentum sums in Eq.~(B11) in Ref.~\cite{Nogga2006} ($S_3,L_3,{\cal L}$)
and obtain a significantly more efficient expression for the $T$-coefficient:
\begin{widetext}
\begin{equation}\label{eq:Tcoef}
\begin{aligned}
\langle N_{\com} L_{\com} N \alpha :JT | pqr:J_{pq}T_{pq}JT\rangle
 &= \delta_{t_{12}T_{pq}} (-1)^{s_{12}+l_{12}+L_{\com}+J_{pq}+j_{3}+3/2}
\sqrt{ [j_{p}][j_{q}][j_{r}][J_{pq}][s_{12}][j_{12}][j_{3}][J_{\rel}] }
 \\& \hspace{-12em} \times
  \sum_{l_{pq}} [l_{pq}]
\left\{
\begin{array}{ccc}
l_{p} & s_{p} & j_{p} \\
l_{q} & s_{q} & j_{q} \\
l_{pq} & s_{12} & J_{pq}
\end{array}
\right\}
\sum_{N_{12}L_{12}} (-1)^{L_{12}} \left\{
\begin{array}{ccc}
L_{12} & l_{12} & l_{pq} \\
s_{12} & J_{pq} & j_{12}
\end{array}
\right\}
\langle N_{12}L_{12},n_{12}l_{12}:l_{pq} | n_{p}l_{p},n_{q}l_{q}:l_{pq} \rangle_{1}
 \\& \hspace{-12em} \times
\sum_{\lambda} (-1)^{\lambda} [\lambda]
\langle N_{\rm cm}L_{\com},n_{3}l_{3}:\lambda | N_{12}L_{12},n_{r}l_{r}:\lambda\rangle_{2}
    \left\{
    \begin{array}{cccccccc}
      j_{12} & & L_{12} & & \lambda & & L_{\rm cm} & \\
             & J_{pq}& & l_{r} & & l_{3} & & J_{\rm rel} \\
      J & & j_{r} & & s_{r} & &j_{3} & \\
    \end{array}
    \right\}.
\end{aligned}
\end{equation}
\end{widetext}
Here, as above, we use the notation $[x]\equiv 2x+1$ and the usual 6-$j$ and 9-$j$ symbols are used.
In addition, we use a
 12-$j$ symbol of the first kind~\cite{khersonskii1988quantum},
 and  $\langle \ldots | \ldots \rangle _{d}$ is Talmi-Moshinsky bracket with mass ratio $d$~\cite{Kamuntavicius2001}.
 For efficiency, 12-$j$ symbols are calculated on the fly from cached 6-$j$ symbols~\cite{khersonskii1988quantum}.
 We have also used $s_{p}=s_{q}=s_{r}=1/2$.
 While \eqref{eq:Tcoef} is a complicated expression, it involves four nested summations (including the expansion of the 12-$j$ symbol; the sum over $N_{12}$ is trivial by energy conservation), rather than the six needed for the expression in Refs.~\cite{Nogga2006,Roth2014}.

Our implementation of the above expressions allows us to generate all the lab-frame three-body matrix elements (with half-precision floating point numbers) needed for a calculation employing the NO2B approximation with a spherical reference state up to $\Etmx=28$ using $\sim 10^{5}$ CPU hours with 187 GB RAM per node.
Importantly, this step only needs to be done once for a given interaction.
Subsequent many-body calculations for different nuclei, or using different methods can be performed using the same file.

\section{\label{sec:conv}Convergence behavior}
Before presenting results for heavy nuclei, we consider the expected convergence behavior of ground-state energies with increasing $\Etmx$.
Knowing the asymptotic behavior enables a controlled extrapolation to $\Etmx\to 3e_{\max}$.
Convergence in $\Etmx$ is distinct from the convergence in $e_{\max}$ (or $N_{\max}$ in the no-core shell model) discussed in Refs.~\cite{Furnstahl2012,Furnstahl2014}, in that the latter deals with a truncation of the Hilbert space, while the former is a truncation on the Hamiltonian.
This means we do not even have an approximate variational principle to rely on.
To simplify the analysis, we assume
that for the soft interactions we consider here, the main contribution to the correlation energy comes from second-order perturbation theory.
Also we assume that $\Etmx$ is sufficiently large that the HF wave function is converged.
The second-order energy correction is
\begin{equation}
     E^{[2]} = \tfrac{1}{4}\sum_{abij}\frac{| \Gamma_{abij}|^2}{\epsilon_i + \epsilon_j - \epsilon_a - \epsilon_b},
\end{equation}
where $\Gamma_{abij}$ contains both two-body and three-body contributions, c.f.~\eqref{eq:GammaM}. Here $i,j$ and $a,b$ run over hole and particle states, respectively.
We can simplify the evaluation by approximating the single-particle energies by the harmonic oscillator energy with the proper energy scale: $\epsilon_{p} = e_{p}\hbar\omega$ with $e_{p}=2n_{p}+l_{p}$ and $\hbar\omega$ the optimal oscillator frequency.
The subscript $p$ indicates either hole or particle state.
We have confirmed that this replacement does not affect the asymptotic behavior.
By increasing the value of $\Etmx$ by one unit the second-order energy changes by
\begin{equation}\label{eq:deltaE2}
    \Delta E^{[2]}
     \approx  \tfrac{1}{2}\sum_{abijk}\frac{V^{\rm NN}_{abij} V^{\rm 3N}_{ijkabk} }{(e_{i}+e_{j}+e_{k}-E_{3\max})\hbar\omega}
    \delta_{\Etmx,e_{a}+e_{b}+e_{k}} \, ,
\end{equation}
where we have assumed $||V^{\rm NN}||\gg ||V^{\rm 3N}||$ and retained only the term linear in $V^{\rm 3N}$.
The interactions we are interested in are regularized by cutoff functions of the form $\exp ( - Q^{2n}/\Lambda^{2n} )$, where $Q$ is a momentum scale, $\Lambda$ the cutoff scale and $n$ some positive power $n$. Depending on the nature of the 3N interaction, $Q$ can be the momentum transfer or the sum of the Jacobi momenta of the form $Q^2 = k_1^2 + 3 k_2^2/4 + k_1'^2 + 3 k_2'^2/4$, where $k_i/k'_i$ are the Jacobi momenta of the initial/final state (see \cite{Hebeler2020} for details).
Then, it is reasonable to assume that the off-diagonal matrix elements are suppressed as:
\begin{equation}\label{eq:Vnnsupp}
V^{\rm NN}_{abij} \approx \bar{V}^{\rm NN}
\exp \left[
- \left( \frac{e_{a}+e_{b}-e_{i}-e_{j}}{\Lambda_{\rm NN}^{2}/m\epsilon_{0}} \right)^{n_{\rm NN}} \right],
\end{equation}
and
\begin{equation}\label{eq:V3nsupp}
V^{\rm 3N}_{abkijk} \approx \bar{V}^{\rm 3N}
\exp \left[
- \left( \frac{ e_{a}+e_{b}+e_{k}-e_{i}-e_{j}-e_{k}
 }{\Lambda_{\rm 3N}^{2}/m\epsilon_{0}} \right)^{n_{\rm 3N}} \right],
\end{equation}
with the cutoff for NN (3N) interaction $\Lambda_{\rm NN}$ ($\Lambda_{\rm 3N}$) and the scale of the NN (3N) interaction $\bar{V}^{\rm NN}$ ($\bar{V}^{\rm 3N}$)
\footnote{Another possible choice would be
\[
V^{\rm NN}_{abij} \approx \bar{V}^{\rm NN}
\exp \left[
- \left( \frac{e_{a}+e_{b}+e_{i}+e_{j}}{\Lambda_{\rm NN}^{2}/m\epsilon_{0}} \right)^{n_{\rm NN}} \right],
\]
and
\[
V^{\rm 3N}_{abkijk} \approx \bar{V}^{\rm 3N}
\exp \left[
- \left( \frac{ e_{a}+e_{b}+e_{k}+e_{i}+e_{j}+e_{k}
 }{\Lambda_{\rm 3N}^{2}/m\epsilon_{0}} \right)^{n_{\rm 3N}} \right].
\]
Even with these forms, one can obtain Eq.~\eqref{eq:extrap} by introducing $X=E_{\rm 3max}+\mu$, $\mu\approx 2e_{\rm F}$ and assuming the condition $E_{\rm 3max} \gg e_{\rm F}$.
}.
For the sake of simplicity, we also assume $n_{\rm NN}=n_{\rm 3N}\equiv n$ in the following.
The above suppression is also found in SRG-evolved potentials~\cite{Jurgenson2008}.
To further simplify, we take the most relevant excitations, i.e., excitations from the Fermi level (with energy $e_{F}$) to the unoccupied orbits.
In this case the numerator in both \eqref{eq:Vnnsupp} and \eqref{eq:V3nsupp} become, when combined with the $\delta$ in \eqref{eq:deltaE2}, equal to $E_{\rm 3max}-3e_F$.
Introducing the new scale factor $1/\sigma^{n} = m^{n}\epsilon^{n}_{0}(1/\Lambda_{\rm NN}^{2n} + 1/\Lambda_{\rm 3N}^{2n})$ and taking the summation explicitly, we obtain the form
\begin{align}\label{eq:delta_e2_approx}
\Delta E^{[2]} \approx
( A_{1} + A_{2}X + A_{3}X^{2}) \exp\left[ -\frac{X^{n}}{\sigma^{n}}\right],
\end{align}
with $X=(E_{\rm 3max} - \mu)$, $\mu \approx 3e_{\rm F}$.
We expect the correction $E^{[2]}$ to be a smooth function of $\Etmx$ in the asymptotic limit, and so we treat the difference $\Delta E^{[2]}$ as a derivative and integrate to obtain the form of $E^{[2]}$:
\begin{equation}
\label{eq:true-form}
E^{[2]} \approx A_{1} \gamma_{\frac{1}{n}} (x)
 +A_{2} \gamma_{\frac{2}{n}}(x)   +
A_{3} \gamma_{\frac{3}{n}}(x)  +
C,
\end{equation}
with $x = [(E_{3\max}-\mu)/\sigma]^{n}$.
Here, $\gamma_{s}(x)$ is the incomplete gamma function:
\begin{equation}
\gamma_{s}(x) = \int^{x}_{0} t^{s-1}e^{-t} dt.
\end{equation}
It turns out that the functions $\gamma_{\frac{1}{n}} (x)$, $\gamma_{\frac{2}{n}} (x)$, $\gamma_{\frac{3}{n}} (x)$ show the same asymptotic behavior, and are therefore redundant for our purposes, so we may simply retain one of the $\gamma$ functions in~\eqref{eq:true-form},
and we choose $\gamma_{\frac{2}{n}}(x)$.
Assuming that the HF energy is well converged with respect to $E_{3\max}$, the formula for the $E_{3\max}$ extrapolation is
\begin{equation}
\label{eq:extrap}
E \sim A \gamma_{\frac{2}{n}} \left[  \left( \frac{E_{3\max}-\mu}{\sigma} \right)^{n} \right] + C.
\end{equation}

It remains to select a reasonable value of the power $n$ entering in \eqref{eq:extrap}.
An SRG-evolved interaction will go as $\exp[-s(k^2-k'^2)^2]$~\cite{Jurgenson2008} with relative momenta $k$ and $k'$, which suggests a value $n=2$.
For the interaction under consideration, 1.8/2.0(EM), the 3N force is not SRG evolved, but instead comes with a regulator $\sim \exp[-(Q^2/\Lambda^2)^4]$~\cite{Hebeler2011}, suggesting $n=4$.
We deal with this ambiguity by checking $n=2,4,6$ to explore the sensitivity to the choice.

Furthermore, from the perturbative expansion of one-body density matrix, we can expect the same $E_{\rm 3max}$ asymptotic behavior for the expectation value of the mean-squared radius operator $\langle r^{2} \rangle$, or any other predominantly one-body operator.

We emphasize that all the discussions are based on the softness of the employed nuclear interaction enabling us to derive the expression through the MBPT.
For a harder interaction, where the MBPT breaks down, we may observe a different convergence pattern with respect to $E_{\rm 3max}$.

\section{\label{sec:results}Numerical results}
The many-body calculation methods used in the following are HF basis many-body perturbation theory (HF-MBPT) and IMSRG.
For open-shell systems, we use the valence-space IMSRG (VS-IMSRG).
For all the many-body methods, we store the usual NN and NO2B 3N matrix elements in RAM, perform the HF calculation to optimize the single-particle basis, and obtain the normal ordered matrix elements~\eqref{eq:E0_J},\eqref{eq:f_J} and \eqref{eq:Gamma_J}.
For open-shell systems, we use an ensemble reference for the normal ordering to capture the 3N interaction effect of valence nucleons as much as possible within the spherical basis framework~\cite{Stroberg2017,Stroberg2019}.
In addition to the HF calculation, we evaluate the correlation energy with MBPT.
Based on a soft nuclear interaction, it was shown that the computationally cheap second- or third-order MBPT can provide results comparable with those from the coupled-cluster method~\cite{Tichai2016,Tichai2018,Tichai2020}.
We could confirm these results in our calculations, and so we use HF-MBPT for the calculations with a large $e_{\max}$ space, where the IMSRG is considerably more expensive.
Our IMSRG calculations are performed with the Magnus formulation~\cite{Morris2015}, using the arctangent generator. Details of the method may be found in recent reviews~\cite{Hergert2016,Stroberg2019}.
The IMSRG and VS-IMSRG calculations are done with the \texttt{imsrg++}~\cite{imsrg++} code, and the subsequent shell-model diagonalizations are done with the \texttt{NuShellX@MSU}~\cite{NuShell} and \texttt{KSHELL}~\cite{KSHELL} codes.

\subsection{$\Etmx$ convergence around $^{132}$Sn}
Here, we investigate large-$\Etmx$ calculations around $^{132}$Sn using the well-established NN+3N interaction 1.8/2.0 (EM)~\cite{Hebeler2011}, which accurately reproduces binding energies to $A\sim 100$~\cite{Simonis2017,Morris2018,Stroberg2021}.
We employ an oscillator basis with frequency $\hbar\omega=16$ MeV, which is near the optimal value giving the most rapid $e_{\rm max}$ convergence for the ground-state energies and radii of the medium-mass nuclei (converged results are independent of $\hbar\omega$)~\cite{Simonis2017}.

One important feature of the 1.8/2.0 (EM) interaction for our purposes is that, while the NN force is softened by a free-space similarity renormalization group (SRG) evolution to a scale $\lambda_{\rm SRG}=1.8~\fm^{-1}$, the corresponding 3N interactions are not SRG evolved. Instead, the cutoff is chosen to be $\Lambda_{\rm 3N}=2.0~\fm^{-1}$ and the short-range low-energy constants $c_D$ and $c_E$ are refit to the triton binding energy and $^{4}$He radius.
This means that we can avoid SRG evolution of the 3N interaction, which introduces additional challenges due to basis truncations (we address these in section \ref{sec:SRG3N}).

\begin{figure}[t]
    \centering
    \includegraphics[width=1.0\columnwidth]{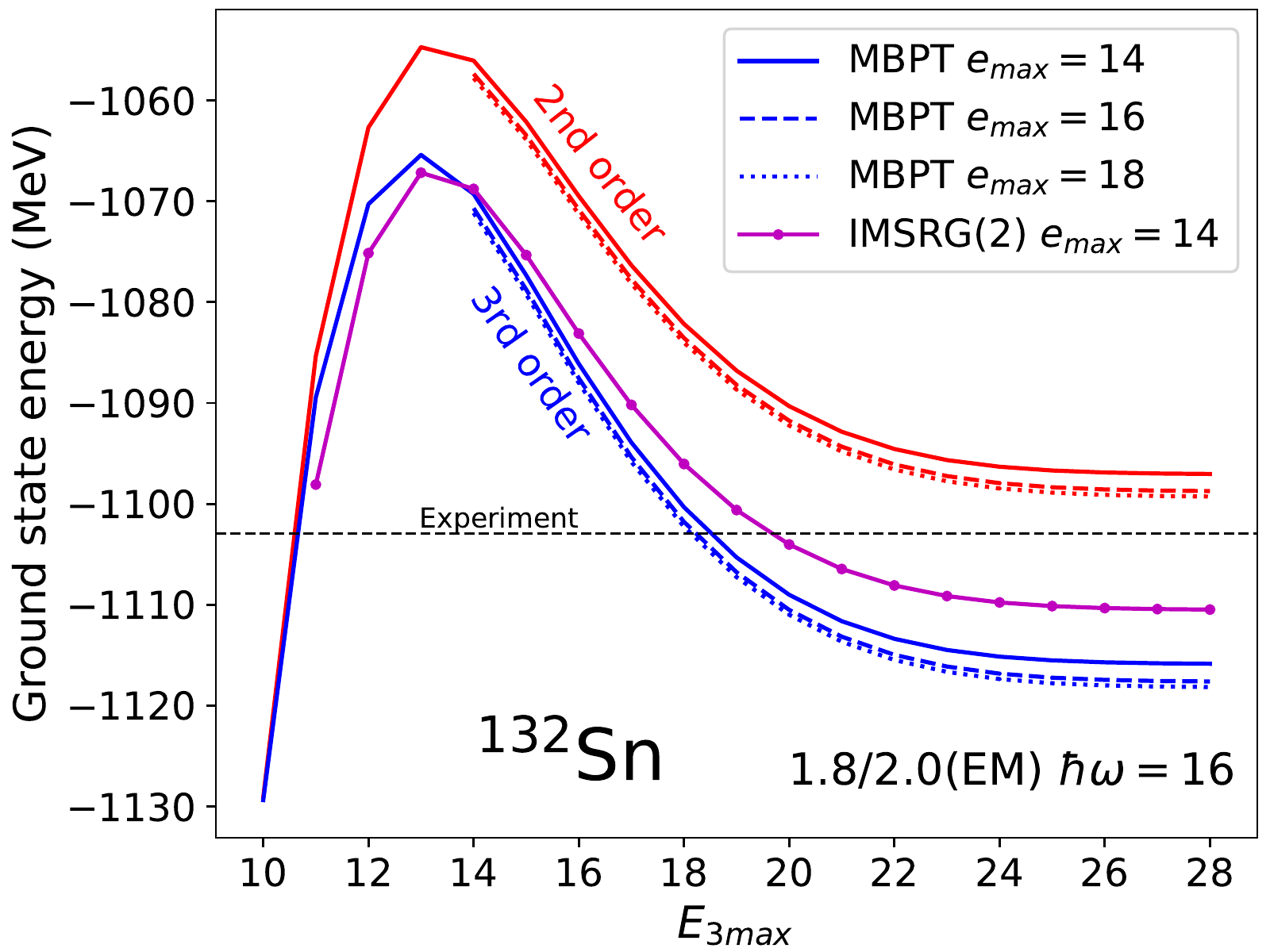}
    \caption{Ground state energy of $^{132}$Sn as a function of $\Etmx$, computed in many-body perturbation theory to second and third order and in IMSRG(2).}
    \label{fig:Sn132MBPT}
\end{figure}

\begin{figure}
    \centering
    \includegraphics[width=\columnwidth]{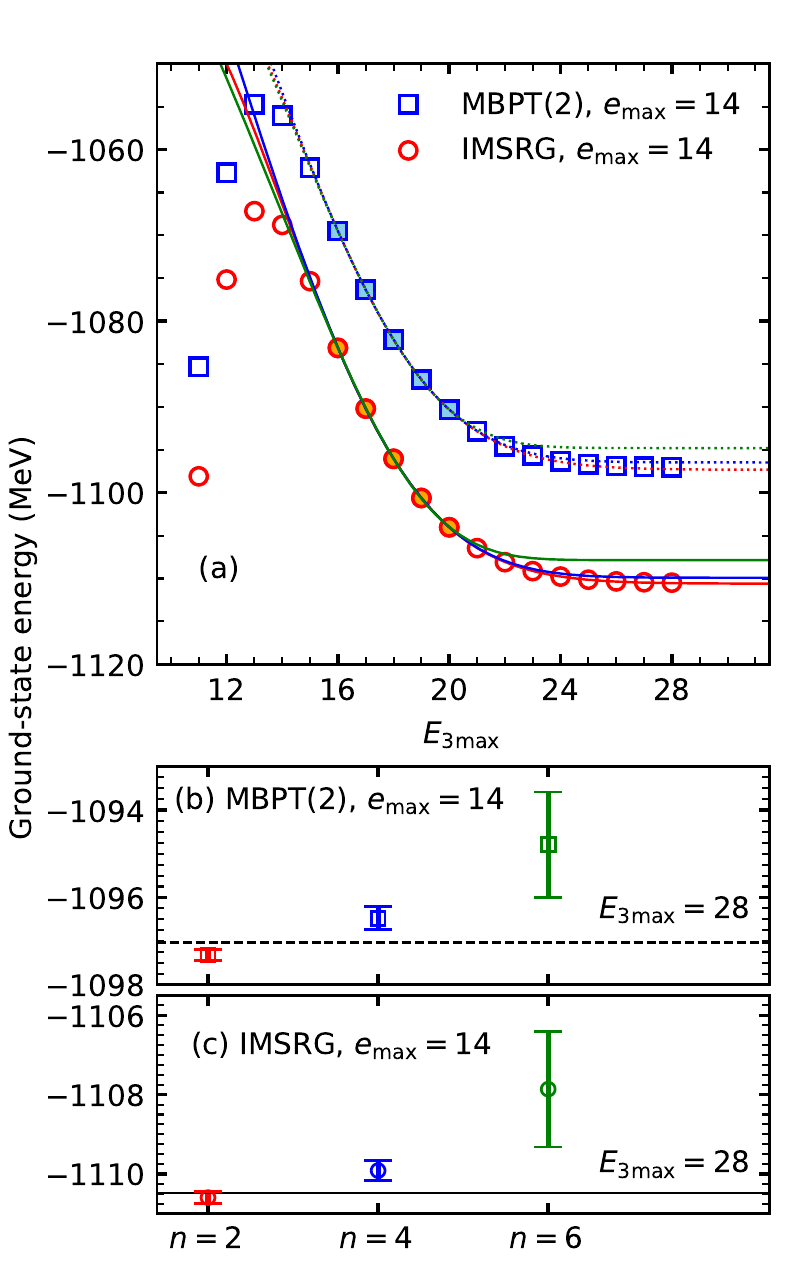}
    \caption{(a) The ground state energy of $^{132}$Sn computed in MBPT(2) and IMSRG(2), as a function of $\Etmx$, and the extrapolated energies for (b) MBPT(2) and (c) IMSRG. The points used in the fitting procedure are indicated by the solid symbols in panel (a).
    The dashed and solid curves are obtained by fitting the functions using $n=2,4,6$ in Eq.~(\ref{eq:extrap}) with the data points of MBPT(2) ($e_{\max}=14$) and IMSRG ($e_{\max}=14$) results, respectively. In panels (b) and (c), the energies are extrapolated to $E_{\rm 3max}=28$.
    The error bars indicate the standard deviation of the distribution, which are obtained with $10^4$ samples drawn from the covariance matrix of the fit.
    }
    \label{fig:E3maxExtrap}
\end{figure}

In Fig.~\ref{fig:Sn132MBPT} we show the ground-state energy of $^{132}$Sn calculated with HF-MBPT and IMSRG as a function of $\Etmx$.
The non-variational nature mentioned above is evident, and is present even at the mean-field level.
We see that truncations at $\Etmx=22$ or $24$ are sufficient to obtain convergence within a few MeV.
For all points in Fig.~\ref{fig:Sn132MBPT}, the 3N matrix elements are stored and read in using half-precision floating point numbers to reduce the memory footprint.
Up to $\Etmx=24$, we can use single-precision numbers to check the impact of this choice.
At $e_{\rm max}=14$, $\Etmx=24$, the half-precision calculation yields HF energies shifted by $-2.14$ MeV, while the second- and third-order MBPT corrections are changed by 0.68~MeV and 0.11~MeV, respectively, yielding a total difference up to third order of $-1.35$ MeV.
This is completely negligible compared with uncertainties arising from many-body truncations (which we expect to be on the order of 20 MeV here\footnote{This estimate is based on the difference between the MBPT(2), MBPT(3) and IMSRG(2) energies, and is consistent with Ref.~\cite{Tichai2016} where the error at MBPT(3) for similarly soft interactions was found to be 0.1-0.2 MeV per particle. We have further corroborated this estimate with MBPT(4) calculations in a smaller $e_{\rm max}$ space.}) and the interaction itself.
We also show in Fig.~\ref{fig:Sn132MBPT} the convergence with respect to $\emax$.
At $\Etmx=28$, the third-order energies for $\emax=14,16,18$, are $-1115.85$~MeV, $-1117.61$~MeV, and $-1118.16$~MeV, respectively, demonstrating convergence at the $1$~MeV level.

Since the second-order correction of $\sim -300$ MeV is much larger than third-order correction of $\sim -20$ MeV, the correlation energy is dominated by second-order correction.
This supports the claim that the extrapolation formula Eq.~(\ref{eq:extrap}) based on the second-order energy correction is applicable in the case of the HF-MBPT(3) and IMSRG, which includes correlations beyond second order.
In panel (a) of Fig.~\ref{fig:E3maxExtrap}, we show $n=2,4,6$ curves of Eq.~(\ref{eq:extrap}) fitted with the HF-MBPT(2) and IMSRG energy results at $e_{\max}=14$, indicated by the solid symbols in the panel.
We see that Eq.~(\ref{eq:extrap}) works for IMSRG energies as well.
Panels (b) and (c) show the extrapolated energies to $\Etmx=28$, which is the largest value we can calculate.
Since the extrapolated point is finite, the uncertainty of all the fitting parameters can propagate to uncertainty of the extrapolated energies.
The uncertainty of the energy is estimated as the standard deviation of the 10000 samples generated with the covariance matrix from the fit.
Comparing the extrapolated and calculated energies, we see that $n=2$ (Gaussian) reproduces the energies for both HF-MBPT(2) and IMSRG cases, and $n=2$ is the most likely to reproduce the convergence behavior in this case.
With $n=2$ formula, we observed that the extrapolated energy to $E_{3\max}=42$ is $-1110.57(2)$ [$-1097.13(2)$] MeV using the IMSRG [HF-MBPT(2)] data $18 \le E_{3\max} \le 23$.

\begin{figure}[t]
    \centering
    \includegraphics[width=1.0\columnwidth]{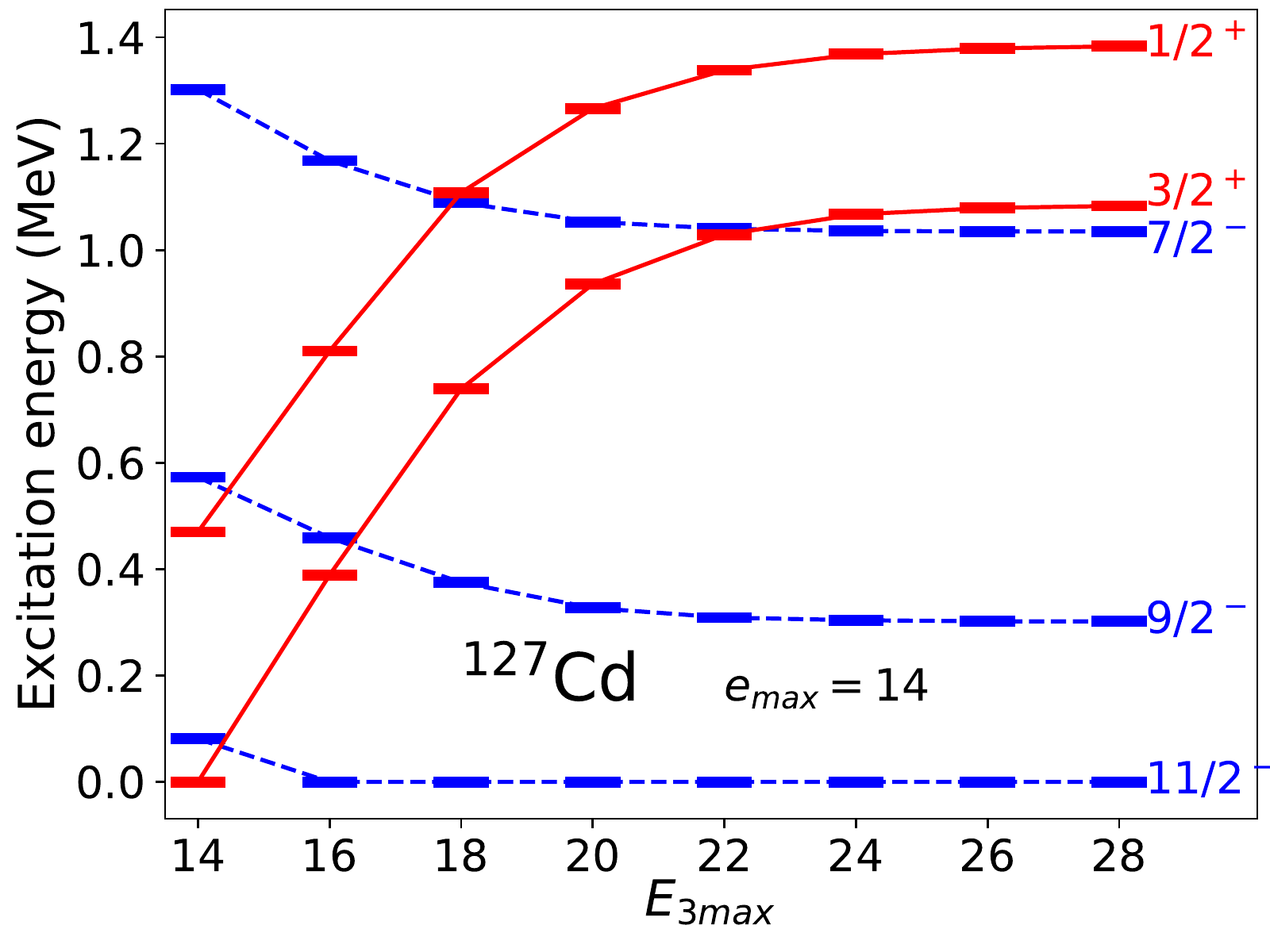}
    \caption{\label{fig:CdSpec}Excitation spectrum of $^{127}$Cd as a function of $\Etmx$, computed in the VS-IMSRG(2) approximation.}
\end{figure}
As already mentioned in Sec.~\ref{sec:intro}, we have observed a lack of convergence with respect to $\Etmx$ in some calculations of heavier systems.
One particular example is $^{127}$Cd as discussed in Ref.~\cite{Lascar2017}.
We revisit the calculations in that work, obtained with the VS-IMSRG, and extend them to larger $\Etmx$.
Here, our single-particle basis truncation is $e_{\max}=14$, and we take the valence space as $\{1p_{3/2}, 1p_{1/2}, 0f_{5/2}, 0g_{9/2}\}$ for protons and $\{1d_{5/2}, 2s_{1/2}, 1d_{3/2}, 0g_{7/2}, 0h_{11/2}\}$ for neutrons above $^{78}$Ni core.
As seen in Fig~\ref{fig:CdSpec}, by $\Etmx=28$ we obtain convergence in excitation energies at the level of 5~keV.
With the previous limit of $\Etmx=18$ there is no sign of convergence.
This behavior can be understood by noting that the $h_{11/2}$ orbit with $e\geq 5$, is impacted by the $\Etmx$ cut differently than the other neutron valence orbits which have $e\geq 4$, and that the parity of a state is driven by the occupation of the $h_{11/2}$.
An analogous argument applies to the proton orbits.

\begin{figure}
    \centering
    \includegraphics[width=1.0\columnwidth]{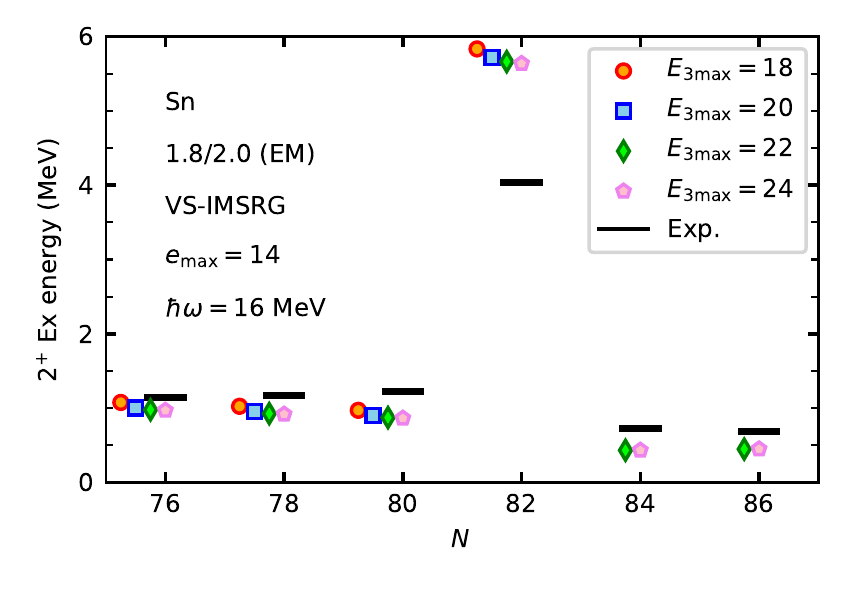}
    \caption{First $2^{+}$ excitation energies of the tin isotopes calculated with the VS-IMSRG(2) approximation.
    The black bars indicate the experimental data~\cite{NationalNuclearDataCenter}.
    }
    \label{fig:2+_energies}
\end{figure}

In contrast, when two states have the same number of oscillator quanta in their naive configurations, we expect that their convergence with respect to $\Etmx$ will be similar and so the energy difference will be less sensitive to the $\Etmx$ truncation.
To illustrate this, we present in the top panel of Fig~\ref{fig:2+_energies} the first $2^{+}$ excitation energies of even-mass tin isotopes, obtained with the VS-IMSRG.
The valence space is $\{1p_{3/2}, 1p_{1/2}, 0f_{5/2}, 0g_{9/2}\}$ for protons and $\{2s_{1/2}, 1d_{3/2}, 0h_{11/2}, 1f_{7/2}\}$ for neutrons above a $^{92}$Ni core, indicated by the open symbols in the figure.
During the IMSRG evolution, the center-of-mass (c.m.) motions are separated with the Gl\"ockner-Lawson prescription~\cite{Gloeckner1974} with the coefficient $\beta=3$, and we observe the stability with respect to $\beta$ (see~\cite{Miyagi2020} for a detailed discussion).
The ground-state energies are converged within approximately 2 MeV.
It is clear from the figure that the $2^{+}$ energies show convergence as $\Etmx$ is increased and $\Etmx=18$ is sufficient to see the systematics of the $2^{+}$ energies.
We note that the $2^{+}$ energy of $^{132}$Sn at $\Etmx=18$ and $24$ differ by 200 keV.
We successfully reproduce the $A$-independent excitation energies of the open-shell nuclei, consistent with the seniority picture.
In fact, the analysis of the calculated wave function of $^{126-130}$Sn reveals that our valence-space wave functions of ground and first $2^{+}$ states are dominated more than $70$\% by the seniority $v=0$ and $v=2$ states, respectively\footnote{While such quantitative details about the wave function will in general depend on the details of the IMSRG transformation, this is a relatively simple way to understand the convergence behavior.}.
The relatively fast convergence of the excitation energies with respect to $\Etmx$ reflects the fact that both the ground and excited states are dominated by configurations with the same occupancies in the oscillator basis.
On the other hand, the excitation at $N=82$ is dominated by a single neutron excitation  $0h_{11/2} \to 1f_{7/2}$.
As these orbits have the same naive number of oscillator quanta, dependence on the $\Etmx$ is still mild.
The predicted excitation energy is about 1.5 MeV above the experimental value, which is attributed to the IMSRG(2) approximation, as seen in earlier works~\cite{Simonis2017,Taniuchi2019}.
Efforts to go beyond the IMSRG(2) approximation are underway~\cite{Heinz2021}.

\begin{figure}
    \centering
    \includegraphics[width=\columnwidth]{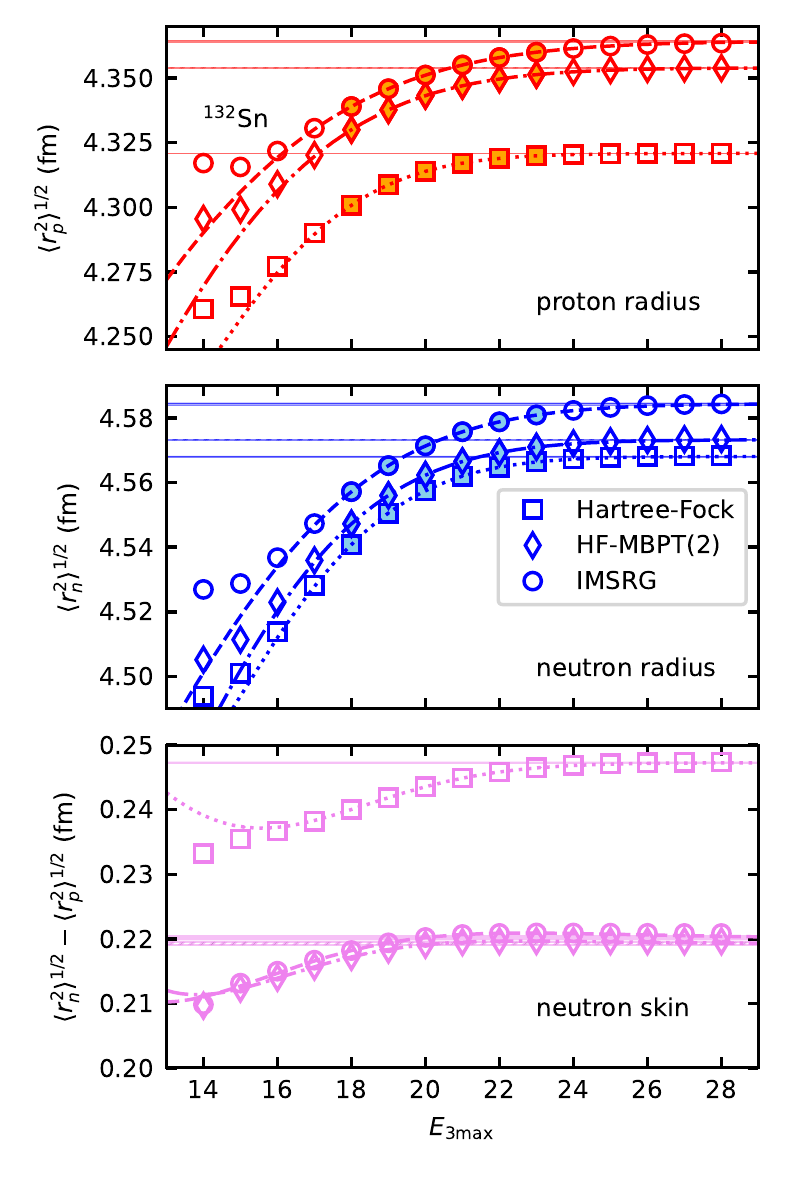}
    \caption{Root-mean-squared point-proton and point-neutron radii, and neutron skin thickness of $^{132}$Sn as a function of $\Etmx$. We use the EM 1.8/2.0 interaction with $\emax=14$ and compute the radii in the Hartree-Fock, HF-MBPT(2), or IMSRG(2) approximations.
    The dotted, dot-dashed, and dashed lines are obtained by fitting to Eq.~(\ref{eq:extrap}).
    The points indicated by the solid symbols are used in the fitting procedure.
    The shaded or hatched bands show the extrapolated radii to $E_{\rm 3max}=3e_{\rm max}=42$ and the widths of the bands are estimated with $10^4$ samples, as in the energy extrapolation.
    }
    \label{fig:Sn132radii}
\end{figure}

Finally, we consider the convergence behavior of point-proton and point-neutron radii through the Hartree-Fock, second-order HF-MBPT, and IMSRG(2).
The diagrams taken into account in the second-order HF-MBPT are listed in Appendix~\ref{appendix:HFMBPT2}.
The charge radii of several isotopes including $^{132}$Sn were recently computed with the self-consistent Green's function method using chiral forces up to $\emax=13$ and $\Etmx=16$~\cite{Arthuis2020}.
We compute point-proton and point-neutron root-mean-squared radii and the neutron skin of $^{132}$Sn as a function of $\Etmx$, and plot the result in Fig.~\ref{fig:Sn132radii}.
We see that convergence is achieved by $\Etmx\sim 22$.
The corresponding converged charge radii are $r_{ch}=4.43$~fm and $4.42$~fm with IMSRG and second-order HF-MBPT, respectively, demonstrating that the effect of the many-body truncation is controllable for radii.
 Eq.~(\ref{eq:extrap}) with $n=2$ reasonably captures the asymptotic convergence behavior of radii.
Also we note that the often-used N$^{2}$LO$_{\rm sat}$ interaction is harder than the interaction employed here, and thus we would expect the calculations with N$^{2}$LO$_{\rm sat}$ will show slower convergence with respect to $E_{\rm 3max}$.
Our converged neutron skin with IMSRG(2) is 0.2202(4)---where the quoted uncertainty only accounts for the $\Etmx$ truncation---consistent with the (model-dependent) extraction~\cite{Klimkiewicz2007} of 0.24(4).

\subsection{SRG evolved NN+3N interaction\label{sec:SRG3N}}
The NN and 3N contributions to the 1.8/2.0 (EM) interaction, used for the calculations discussed in the previous section, are defined at different cutoff and resolution scales.
For a more systematic convergence study of the calculations it would be desirable explore the resolution-scale and cutoff dependence of observables. Using interactions with a higher cutoff, observables in heavy nuclei will be impossible to converge in the largest feasible model spaces, even with the advances discussed in this work. Therefore, these interactions first need to be softened via a free-space SRG evolution~\cite{Bogner2007} (or some other procedure~\cite{Bogner2003,Navratil1998,Feldmeier1998}).
For the following calculations we evolve NN and 3N sectors consistently in the harmonic-oscillator basis space. For the NN sector, the evolution is done within the space spanned by the principle quantum number of the relative motion up to 200.
Assuming our typical basis frequency of a few tens MeV, the UV scale of this space is a few GeV$/c$---sufficiently larger than the typical momentum scale of $\sim 500$ MeV$/c$ of the bare NN interaction from the chiral EFT, and we can safely evolve the NN Hamiltonian.

For the 3N sector, we evolve the 3N Hamiltonian within the space defined by the three-body principle quantum number $N_{3\max}$, the sum of the principle quantum numbers of the motions for corresponding Jacobi variables.
Since the 3N evolution is computationally demanding compared to the NN evolution, we cannot handle a value of $N_{3\max}$ well beyond the typical nuclear interaction scale.
We therefore need to investigate the $N_{3\max}$ dependence as we move to heavier systems, as done in Ref.~\cite{Binder2014}.
In the following, we use the chiral N$^3$LO NN interaction from Entem and Machleidt~\cite{Entem2003} and the N$^2$LO 3N interaction with both local and non-local regulators developed in Ref.~\cite{Soma2019} denoted as NN+3N(lnl).

\begin{figure}[t]
    \centering
    \includegraphics[width=1.0\columnwidth]{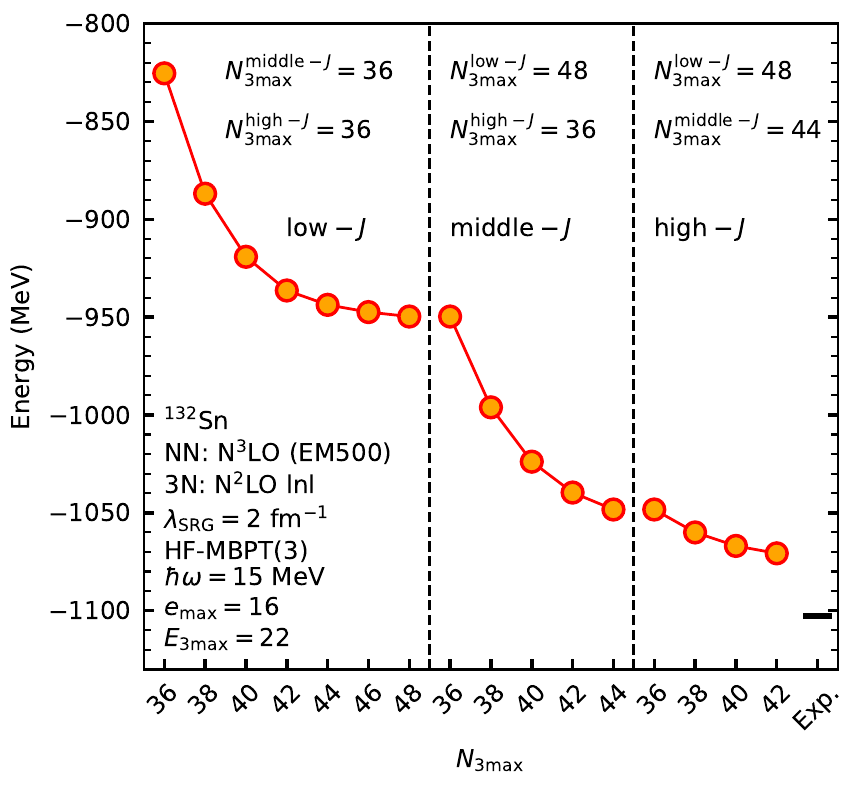}
    \caption{Ground state energy of $^{132}$Sn as a function of $N_{3\max}$ for the SRG evolution, computed in third-order HF basis many-body perturbation theory HF-MBPT(3) at $\hbar\omega=15$ MeV, $e_{\max}=16$, and $E_{3\max}=22$.
    The vertical dashed lines indicate the partitions of
    low-$J$ ($J_{\rm rel} \leq 13/2$), middle-$J$ ($15/2 \leq J_{\rm rel} \leq 21/2$), and high-$J$ ($J_{\rm rel} \geq 23/2$) regions.
    }
    \label{fig:Sn132SRG}
\end{figure}

In Fig.~\ref{fig:Sn132SRG} we show the ground-state energy of $^{132}$Sn as a function of $N_{\rm 3max}$.
Because the Hamiltonian is block diagonal in the relative angular momentum $J_{\rm rel}$, we can apply a different $N_{\rm 3max}$ cut to each $J_{\rm rel}$ block.
We include all channels up to $J_{\rm rel}\leq \Etmx +3/2=47/2$, which is the highest value that can contribute.
To simplify the analysis, we divide the $J_{\rm rel}$ blocks into low-$J$ ($J_{\rm rel}\leq 13/2$), middle-$J$ ($15/2 \leq J_{\rm rel} \leq 21/2$), and high-$J$ ($J_{\rm rel} \geq 23/2$) partitions, and vary $N_{\rm 3max}$ for each partition.
The SRG evolution is run to a scale of $\lambda_{\rm SRG}=2.0$ fm$^{-1}$, working with a basis frequency $\hbar\omega=30$ MeV.
After the evolution, the frequency is converted to $\hbar\omega=15$ MeV for the many-body calculations (see Ref.~\cite{Roth2014} for details).
The many-body calculations are done with third-order HF-MBPT with lab-frame truncations $\emax=16$ and $\Etmx=22$.
In Fig.~\ref{fig:Sn132SRG} we see that
 the low-$J$ and high-$J$ partitions are converged within the level of a few MeV by $N_{\rm 3max}=48$ and $N_{\rm 3max}=42$, respectively, while the middle-$J$ region converges more slowly.
To our knowledge, these are the largest $N_{\rm 3max}$ spaces explored in the literature.
It appears that SRG evolution of the $J_{\rm rel}\gtrsim 15/2$ blocks such that heavy nuclei are converged with respect to $N_{\rm 3max}$ will not be possible in the near term without further technical developments.

\begin{figure}[t]
    \centering
    \includegraphics[width=1.0\columnwidth]{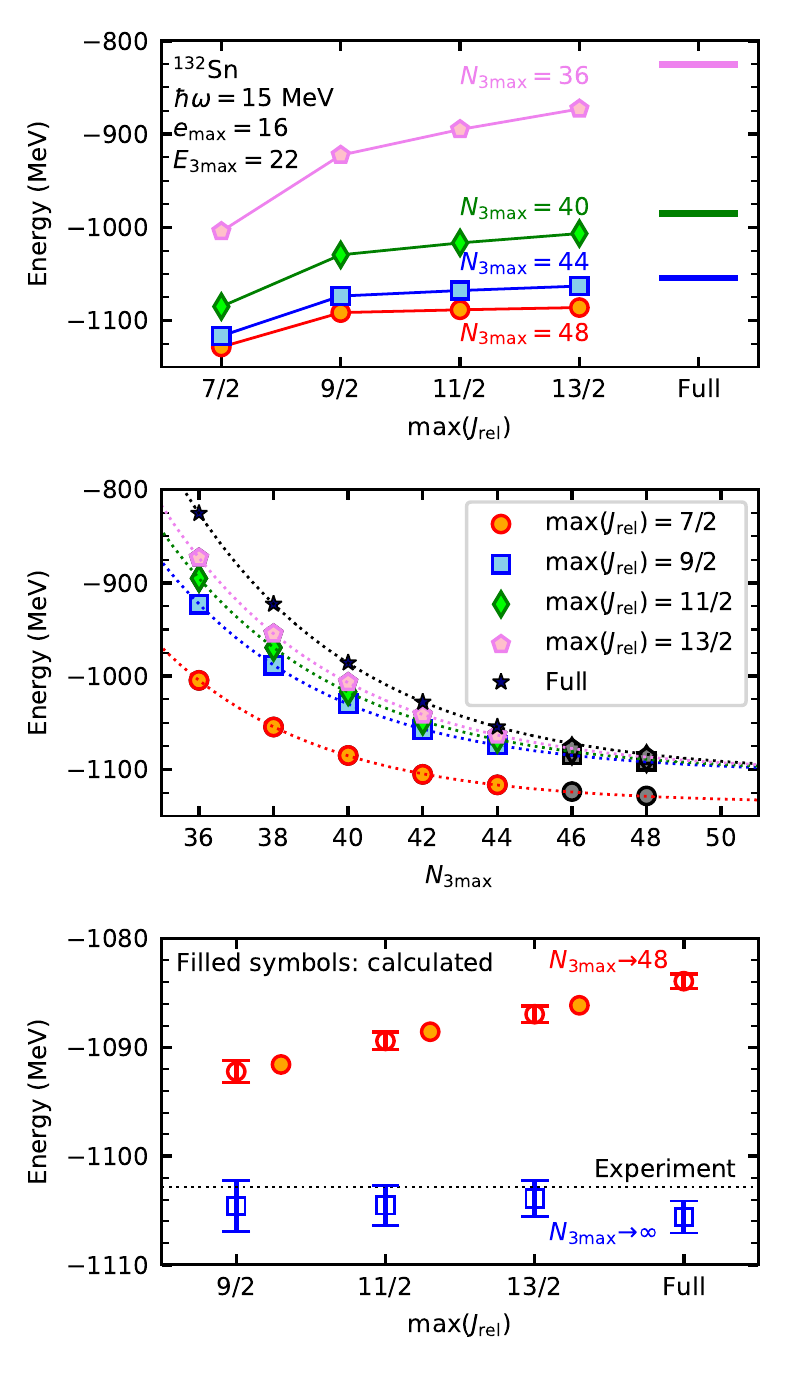}
    \caption{Ground state energy of $^{132}$Sn computed in HF-MBPT(3) at $\hbar\omega=15$ MeV, $e_{\max}=16$, and $E_{3\max}=22$.
    In the top panel, the energies are shown as a function of maximum $J_{\rm rel}$ for the transformation to the lab-frame.
    The middle panel shows the extrapolation of $N_{\rm 3max}$ to the infinity.
    The extrapolated energies are shown as a function of maximum $J_{\rm rel}$ in the bottom panel.
    }
    \label{fig:Sn132SRG_Jmax}
\end{figure}

An alternative truncation scheme we can explore is the maximum value of $J_{\rm rel}$.
Since the nuclear interaction is short range, we naively expect that the high $J_{\rm rel}$ components are suppressed by the angular momentum barrier.
In the top panel of Fig.~\ref{fig:Sn132SRG_Jmax}, the ground-state energy of $^{132}$Sn is shown as a function of the maximum $J_{\rm rel}$ included in the transformation Eq.~(\ref{eq:3Ntrans}).
The points labeled ``Full'' use a uniform $N_{\rm 3max}$ for all blocks up to $J_{\rm rel}=47/2$.
Again, energies are computed with HF-MBPT(3) at $\hbar\omega=15$ MeV, $e_{\max}=16$, and $E_{\rm 3max}=22$.
We observe that as we increase $N_{\rm 3max}$, the contribution of channels with $J_{\rm rel}>9/2$ becomes essentially negligible.

In order to extrapolate to $N_{\rm 3max}\to\infty$, we fit the calculated energies with an exponential function $E = a \exp( -b N_{\rm 3max}) + E_{\infty}$ as shown in the middle panel.
In the fitting procedure, we used the energies at $N_{\rm 3max}=36,38,40,42,44$, and used the $N_{\rm 3max}=48$ points to validate the assumed functional form.\footnote{We also tried fitting with the Gaussian function $E=a \exp(-b N_{\rm 3max}^{2}) + E_{\infty}$, and found this does not provide consistent results with the computed $N_{\rm 3max}=48$ energies.}
The $N_{\rm 3max}$ extrapolated energies are shown in the bottom panel.
The agreement of calculated and extrapolated energies at $N_{\rm 3max}=48$ validates the fitting formula employed here.
The final energy obtained by extrapolating the "full" results to $N_{\rm 3max} \to \infty$ is $-1105.6(15)$~MeV, which agrees within the error bars with the extrapolated energies from $\max(J_{\rm rel})=9/2, 11/2, 13/2$ results.
This reinforces the observation that the contributions from channels with $J_{\rm rel}>9/2$ are negligible in the $N_{\rm 3max}\to \infty$ limit.

\begin{figure}
    \centering
    \includegraphics[width=\columnwidth]{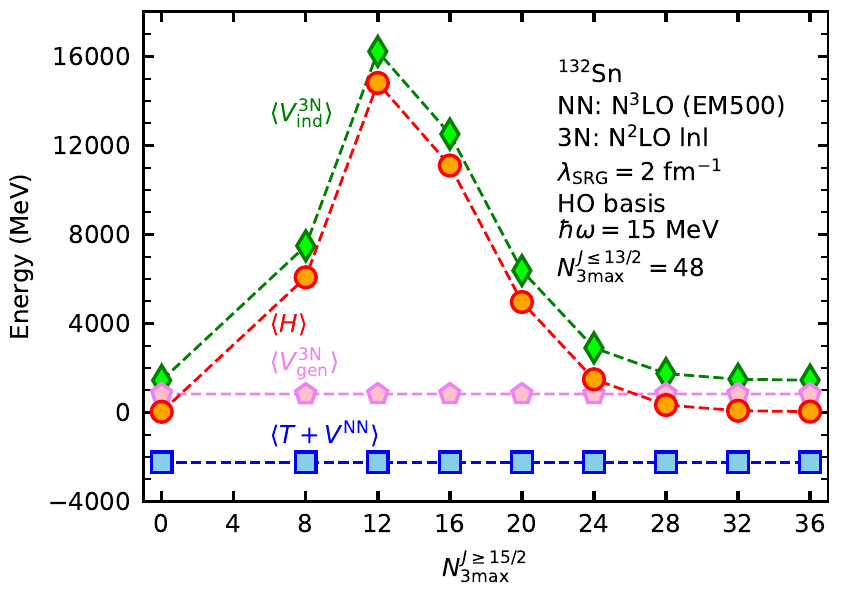}
    \caption{\label{fig:Sn132highJ}
    Harmonic-oscillator basis energy components of $^{132}$Sn as a function of the cut $N_{3\max}$ applied to the SRG basis for states with $J_{\rel}\geq 15/2$.
    For details regarding each component, see main text.
    }
\end{figure}

This result is somewhat surprising as it suggests that we can obtain a more accurate result by neglecting the high-$J_{\rel}$ sector altogether than we can by evolving it in the largest space we can manage.
Evidently, the main impact of the high-$J_{\rel}$ matrix elements is to introduce an artifact due to the $N_{\rm 3max}$ truncation, which is removed in the limit $N_{\rm 3max}\to \infty$.
To investigate the origin of this artifact, in Fig.~\ref{fig:Sn132highJ} we hold fixed $N_{\rm 3max}=48$ for the $J_{\rel}\leq 13/2$ partition, and plot the expectation values $\langle T+V^{\rm NN}\rangle$, $\langle V^{\rm 3N}_{\rm ind} \rangle$, $\langle V^{\rm 3N}_{\rm gen}\rangle$, and $\langle H\rangle$ as a function of the $N_{3\max}$ cut applied to the $J_{\rel}\geq 15/2$ partition.
Here $T$ is the relative kinetic energy, $V^{\rm NN}$ is the evolved NN potential, $V^{\rm 3N}_{\rm ind}$ is the induced 3N potential, $V^{\rm 3N}_{\rm gen}$ is the evolved ``genuine'' 3N potential, and $H$ is the transformed Hamiltonian obtained by summing all of the kinetic and potential terms.
The expectation values are taken in a naive harmonic oscillator ground state of $^{132}$Sn.

At $N_{\rm 3max}=0$, corresponding to the $J=13/2$ point in Fig.~\ref{fig:Sn132SRG_Jmax}, we obtain a bound energy.
With increasing $N_{\rm 3max}$ the energy shoots up to 15~GeV, driven by the $\langle V^{\rm 3N}_{\rm ind} \rangle$ component, before converging back towards the $N_{\rm 3max}=0$ value.
It appears that the impact the high-$J_{\rm rel}$ matrix elements are negligible.
Similar behavior is found in $^{78}$Ni, where the fully-converged and $N_{\rm 3max}=0$ HF energies differ by 0.3 MeV.

To further investigate the enormous contributions from induced 3N interactions, we decompose $\langle V^{\rm 3N}_{\rm ind}\rangle$ into terms induced by transforming the one-pion exchange, two-pion exchange, and contact parts of $V_{\rm NN}$, as well as the kinetic energy.
We find that at $N_{\rm 3max}=16$, all four of these induced terms contribute several GeV to the energy in Fig.~\ref{fig:Sn132highJ}, indicating that this behavior is generic and not tied to the detailed structure of the NN interaction.
Further understanding of the mechanism of this large induced component should be pursued, as it may point the way to a more efficient treatment.

Through this analysis, we conclude that one can perform a more accurate 3N SRG evolution with a truncation in $J_{\rm rel}$, rather than using all possible $J_{\rm rel}$ channels without fully achieving the convergence with respect to $N_{\rm 3max}$.
We leave for future work the question of whether this holds for other operators.

\begin{figure}
    \centering
    \includegraphics[width=\columnwidth]{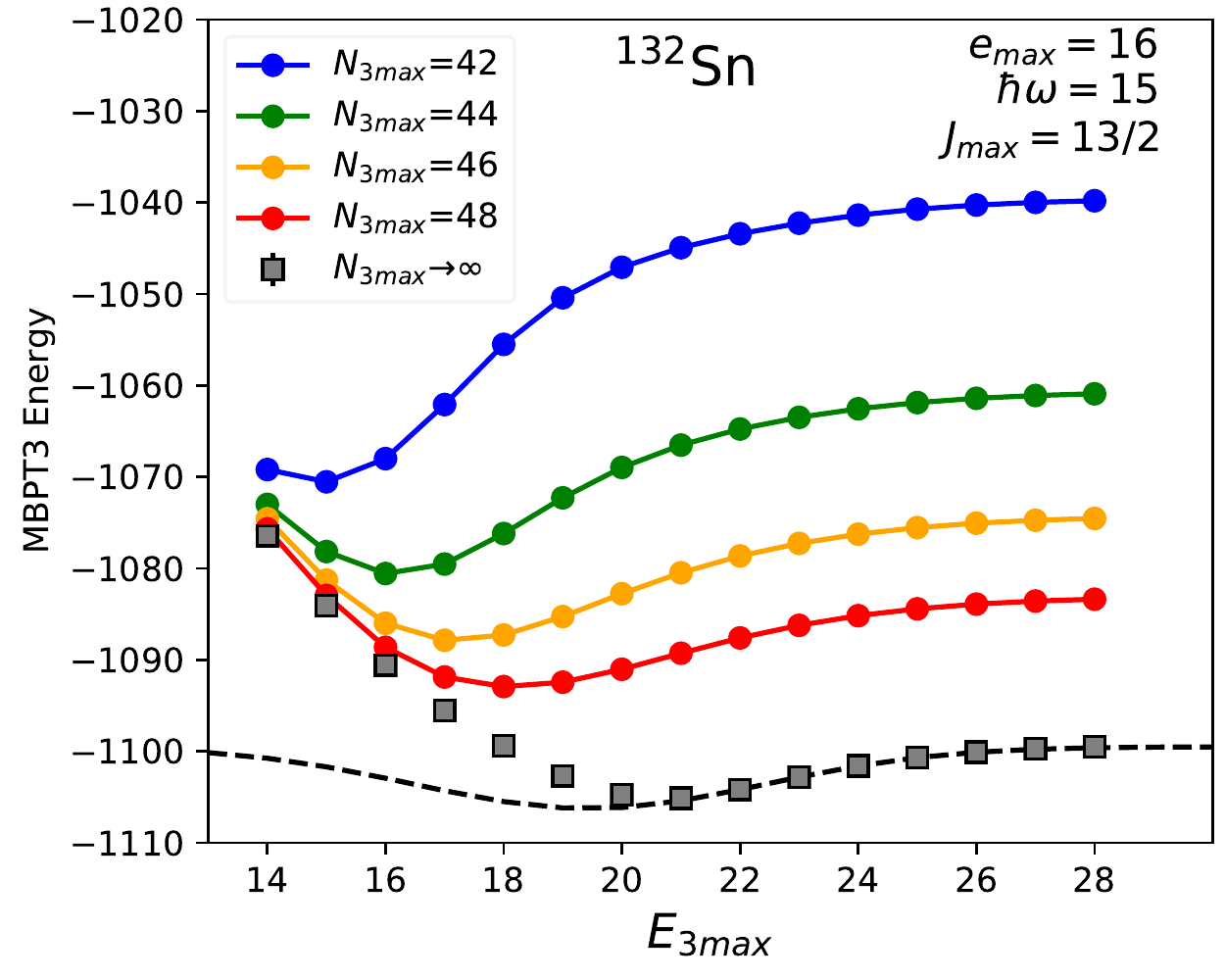}
    \caption{Ground state energy of $^{132}$Sn obtained in 3rd order HF-MBPT, as a function of $\Etmx$ for several values of $N_{\rm 3max}$.
    We retain relative angular momenta $J_{\rel} \leq 13/2$ in the transformation to the single-particle coordinate.
    The gray squares are extrapolated to $N_{\rm 3max}\to \infty$ with an exponential, and then fit with a Gaussian (dashed line) to extrapolate in $\Etmx$.}
    \label{fig:E3maxN3maxConv}
\end{figure}

Finally, we demonstrate that the asymptotic convergence in $\Etmx$ discussed in section~\ref{sec:conv} is also observed for a consistently SRG-evolved NN+3N interaction.
In Fig.~\ref{fig:E3maxN3maxConv}, we show the 3rd order HF-MBPT ground-state energy of $^{132}$Sn as a function of $\Etmx$, at multiple values of $N_{\rm 3 max}$ for the case $J_{\rm rel}\leq 13/2$ (similar behavior is observed for $J_{\rm rel}\leq 9/2$).
In contrast to the unevolved case, we observe an increase in the energy for large $\Etmx$.
This bump diminishes with increasing $N_{\rm 3max}$, indicating that the truncation artifact shows up most significantly in the large $\Etmx$ matrix elements, as would be expected.
For each $\Etmx$, we extrapolate to $N_{\rm 3max}\to\infty$ using an exponential form, and we obtain the gray squares in Fig.~\ref{fig:E3maxN3maxConv} (the extrapolation uncertainties are smaller than the markers).

The extrapolated points still display a minimum as a function of $\Etmx$ before converging to the final answer from below.
The decreasing trend below $\Etmx=20$ is driven by the convergence of the HF energy, while the increase above $\Etmx=20$ is driven by the second order MBPT correction.
The fact that the energy converges from below in this case supports the assumption in section~\ref{sec:conv} that the asymptotic convergence in $\Etmx$ is driven by the $V_{NN}$-$V_{3N}$ cross-term , which can be either positive or negative.
The asymptotic behavior is fit well with a Gaussian with similar parameters (aside from the overal sign) to those in the unevolved case.

The extrapolated ground-state energy for $^{132}$Sn is then $-1099.502(3)$ MeV, where this tiny uncertainty only accounts for the fit uncertainty in the $\Etmx$ and $N_{\rm 3max}$ extrapolations.
This uncertainty is clearly negligible compared with the many-body uncertainty (we only use third-order MBPT), the $e_{\rm max}$ truncation uncertainty, effects of induced 4N forces, contributions from higher orders in the EFT expansion, and the fact that we use a half-precision floating point representation for storing the 3N matrix elements.
We note that the effect of the SRG induced many-body interactions can be accessed by checking the $\lambda_{\rm SRG}$ dependence.
Our almost-converged $^{132}$Sn calculations at $e_{\rm max}=16$ and $E_{\rm 3max}=22$ show that the ground-state energy changes about 20 MeV within $\lambda_{\rm SRG}=1.8-2.2$ fm$^{-1}$ range, which is at the 2\% level of the total ground-state energy.
As the other sources of uncertainty likely contribute at the level of a few tens of MeV, the NN+3N(lnl) interaction is in excellent agreement with the experimental value of $-1102.8$ MeV~\cite{Wang2017}, especially considering that it was fit to the properties of $A\leq 4$ nuclei.

\section{\label{sec:conclusion}Conclusion}
In this work we introduce a framework in which only 3N matrix elements relevant for the NO2B approximation are stored in memory, which reduces the memory requirement by approximately two orders of magnitude.
This enables us to generate lab-frame 3N matrix elements up to $\Etmx=28$, significantly larger than the previous limit of $\Etmx=18$.
We further discussed the asymptotic behavior of the ground-state energy with respect to the $E_{\rm 3max}$ truncation, which allows controlled extrapolations to $\Etmx=3e_{\max}$.
To explore the applicability of the ab initio calculation, we empolyed the HF-MBPT(2), HF-MBPT(3), and IMSRG(2) to solve the many-body Schr\"odinger equation.
Using the established 1.8/2.0 (EM) interaction, we obtained the ground-state energies converged at the level of 1 MeV (with respect to the $\emax$ and $\Etmx$ truncations) around $^{132}$Sn.
As illustrated in the $^{127}$Cd case, convergence in $\Etmx$ is essential not just for ground states but for spectroscopy as well.
Even with this substantially larger lab-frame $E_{\rm 3max}$ cut, as we move to the heavy-mass region, convergence with respect to truncations made in the free-space 3N SRG evolution pose an additional challenge.
Using the N$^{3}$LO NN + N$^{2}$LO 3N(lnl)~\cite{Soma2019} interaction, we have demonstrated that a truncation $J_{\rel}~\lesssim 13/2$ is more accurate (not to mention less costly) for calculations of ground state energies than including larger $J_{\rel}$, if full convergence in those channels cannot be achieved. A corresponding convergence analysis for excited states and other observables with respect to $J_{\rel}$ remains future work.

This work lifts the primary limitation that has thus far kept ab initio calculations constrained to the $A\lesssim 100$ region.
Among the studies that will be enabled are the neutron skin of $^{208}$Pb~\cite{Piekarewicz2019}, neutrinoless double-beta decays and dark matter searches in germanium and selenium~\cite{Belley2020}, as well as xenon~\cite{Gando2016} and tellurium~\cite{Alfonso2015}, and investigations of nuclear matter parameters based on the response functions of heavy nuclei~\cite{Hu2021}.

\begin{acknowledgments}
We thank A.~Bansal, H.~Hergert, K.~Kravvaris, A.~McCoy, R.~Roth, J.~Simonis, R.~Wirth and J.~M.~Yao for valuable discussions. TRIUMF receives funding via a contribution through the National Research Council of Canada.
SRS is supported by the U.~S. Department of Energy under Contract DE-FG02-97ER41014. PN acknowledges support from the NSERC Grant No. SAPIN-2016-00033 and JDH from NSERC Grants SAPIN-2018-00027 and RGPAS-2018-522453. KH acknowledges support from the Deutsche Forschungsgemeinschaft (DFG, German Research Foundation) -- Project-ID 279384907 -- SFB 1245.
Computations were performed with an allocation of computing resources on Cedar at WestGrid and Compute Canada, and on the Oak Cluster at TRIUMF managed by the University of British Columbia department of Advanced Research Computing (ARC).

\end{acknowledgments}

\appendix

\section{Normal-ordered matrix elements\label{appendix:NO}}
In an uncoupled basis, the expressions for the normal-ordered matrix elements are
\begin{equation}\label{eq:uncoupledE0}
\begin{aligned}
    E_0 = &\sum_{p'p} \rho_{p'p}t_{p'p} + \frac{1}{4}\sum_{pp'qq'} \rho_{p'q'pq} V^{\rm NN}_{p'q'pq}\\
    &+\frac{1}{36}\sum_{pp'qq'rr'} \rho_{p'q'r'pqr}V^{\rm 3N}_{p'q'r'pqr}
    \end{aligned}
\end{equation}
\begin{equation}\label{eq:uncoupledf}
\begin{aligned}
    f_{p'p} = &t_{p'p} + \sum_{q'q}\rho_{q'q}V^{\rm NN}_{p'q'pq}\\
    &+ \frac{1}{4}\sum_{qq'rr'}\rho_{q'r'pr}V^{\rm 3N}_{q'r'p'qrp}
    \end{aligned}
\end{equation}
\begin{equation}\label{eq:uncoupledGamma}
    \Gamma_{p'q'pq} = V^{\rm NN}_{p'q'pq} + \sum_{rr'}\rho_{r'r}V^{\rm 3N}_{p'q'r'pqr}.
\end{equation}
In \eqref{eq:uncoupledE0} \eqref{eq:uncoupledf}, \eqref{eq:uncoupledGamma}, we have used the density matrices
\begin{equation}
    \begin{aligned}
        \rho_{p'p} &\equiv \langle \Phi | a^{\dagger}_{p'}a_{p} |\Phi \rangle \\
        \rho_{p'q'pq} &\equiv \langle \Phi | a^{\dagger}_{p'}a^{\dagger}_{q'}a_{q}a_{p} | \Phi \rangle\\
        \rho_{p'q'r'pqr} &\equiv \langle \Phi | a^{\dagger}_{p'}a^{\dagger}_{q'}a^{\dagger}_{r'}a_{r}a_{q}a_{p} | \Phi \rangle\\
    \end{aligned}
\end{equation}
taken for some general reference state $|\Phi\rangle$.
If the reference $|\Phi\rangle$ is spherically symmetric, then the density matrices may be expressed in a $J$-coupled form defined by
\begin{align}
    \rho_{p'p} &= \rho_{p'p} \delta_{j_{p'}j_{p}}\delta_{m_{p'}m_{p}} \label{eq:rhodeltas}\\
    \rho_{p'q'pq} &= \sum_{J}
    \mathcal{C}^{ j_{p'}j_{q'}J}_{m_{p'}m_{q'}M}
    \mathcal{C}^{ j_{p}j_{q}J}_{m_{p}m_{q}M}
    \rho_{p'q'pq}^{J}\\
    \rho_{p'q'r'pqr} &= \sum_{J_{pq}J_{p'q'}J}
    \mathcal{C}^{ j_{p'}j_{q'}J_{p'q'}}_{m_{p'}m_{q'}M_{p'q'}}
    \mathcal{C}^{ j_{p}j_{q}J_{pq}}_{m_{p}m_{q}M_{pq}}\\
    &\hspace{1em}\times \mathcal{C}^{ J_{p'q'}j_{r'}J}_{M_{p'q'}m_{r'}M}
    \mathcal{C}^{ J_{pq}j_{r}J}_{M_{pq}m_{r}M}
    \rho_{p'q'r'pqr}^{J_{p'q'}J_{pq}J}\nonumber
    \end{align}
    where the $\mathcal{C}$ are Clebsch-Gordan coefficients.
If $|\Phi\rangle$ does not mix proton and neutron orbits, then \eqref{eq:rhodeltas} will contain an additional $\delta_{t_{zp'},t_{zp}}$.
If $|\Phi\rangle$ has good parity, we also have $\delta_{l_{p'}l_p}$.
For a spherical reference, the expression for the normal-ordered matrix elements becomes
\begin{equation}
 \begin{aligned}
    E_0 &= \sum_{p'p}\rho_{p'p}t_{p'p}
     + \frac{1}{4}\sum_{pp'qq'}\sum_{J}[J] \rho^{J}_{p'q'pq} V^{J}_{p'q'pq} \\
     &+ \frac{1}{36} \sum_{pp'qq'rr'}\sum_{J_{pq}J}[J] \rho^{J_{pq}J_{pq}J}_{p'q'r'pqr} V^{J_{pq}J_{pq}J}_{p'q'r'pqr}
     \end{aligned}
\end{equation}
\begin{equation}
\begin{aligned}
    f_{p'p} &= t_{p'p} + \sum_{q'q}\sum_{J}\frac{[J]}{[j_{p}]}\rho_{q'q}V^{J}_{p'q'pq} \\
    &+ \frac{1}{4}\sum_{qq'rr'}\sum_{J_{qr}J}\frac{[J]}{[j_{p}]} \rho^{J_{qr}}_{q'r'qr}V^{J_{qr}J_{qr}J}_{q'r'p'qrp}
    \end{aligned}
\end{equation}
\begin{equation}\label{eq:Gamma_J_corr}
    \Gamma^{J_{pq}}_{p'q'pq} = V^{J_{pq}}_{p'q'pq}
    + \sum_{r'r J} \frac{[J]}{[J_{pq}]}\rho_{r'r} V^{J_{pq}J_{pq}J}_{p'q'r'pqr}.
\end{equation}
Where we have used un-normalized $J$-coupled matrix elements.
Finally, in the case where $|\Phi\rangle$ is uncorrelated so that $\rho_{p'q'pq}$ and $\rho_{p'q'r'pqr}$ are given by antisymmetrized products of one-body densities (again using the index permutation operators $P$)
\begin{equation}
\begin{aligned}
\rho_{p'q'pq} &= (1-P_{pq}) \rho_{p'p}\rho_{q'q}\\
\rho_{p'q'r'pqr} &= (1-P_{qr})(1-P_{pq}-P_{pr}) \rho_{p'p}\rho_{q'q}\rho_{r'r},
\end{aligned}
\end{equation}
then the normal ordered matrix elements become
\begin{equation}\label{eq:E0_J}
 \begin{aligned}
    E_0 &= \sum_{pp'}\rho_{p'p}t_{p'p}
     + \frac{1}{2}\sum_{pp'qq'} \rho_{p'p}\rho_{q'q} \sum_{J}[J]  V^{J}_{p'q'pq} \\
     &+ \frac{1}{6} \sum_{\substack{pqr\\p'q'r'}} \rho_{p'p}\rho_{q'q}\rho_{r'r} \sum_{J_{pq}J}[J]  V^{J_{pq}J_{pq}J}_{p'q'r'pqr}
     \end{aligned}
\end{equation}
\begin{equation}\label{eq:f_J}
\begin{aligned}
    f_{p'p} &= t_{p'p} + \sum_{qq'} \rho_{q'q} \sum_{J}\frac{[J]}{[j_{p}]}V^{J}_{p'q'pq} \\
    &+ \frac{1}{2}\sum_{qq'rr'} \rho_{q'q}\rho_{r'r} \sum_{J_{qr}J}\frac{[J]}{[j_{p}]} V^{J_{qr}J_{qr}J}_{q'r'p'qrp}
    \end{aligned}
\end{equation}
\begin{equation}\label{eq:Gamma_J}
    \Gamma^{J_{pq}}_{p'q'pq} = V^{J_{pq}}_{p'q'pq}
    + \sum_{rr'} \rho_{r'r}\sum_{J} \frac{[J]}{[J_{pq}]} V^{J_{pq}J_{pq}J}_{p'q'r'pqr}
\end{equation}
where we have omitted the $\delta$s implied by \ref{eq:rhodeltas}.

\section{Ground-state expectation value of a scalar operator in second-order HF-MBPT\label{appendix:HFMBPT2}}
\begin{figure}[t]
    \centering
    \includegraphics[width=1.0\columnwidth]{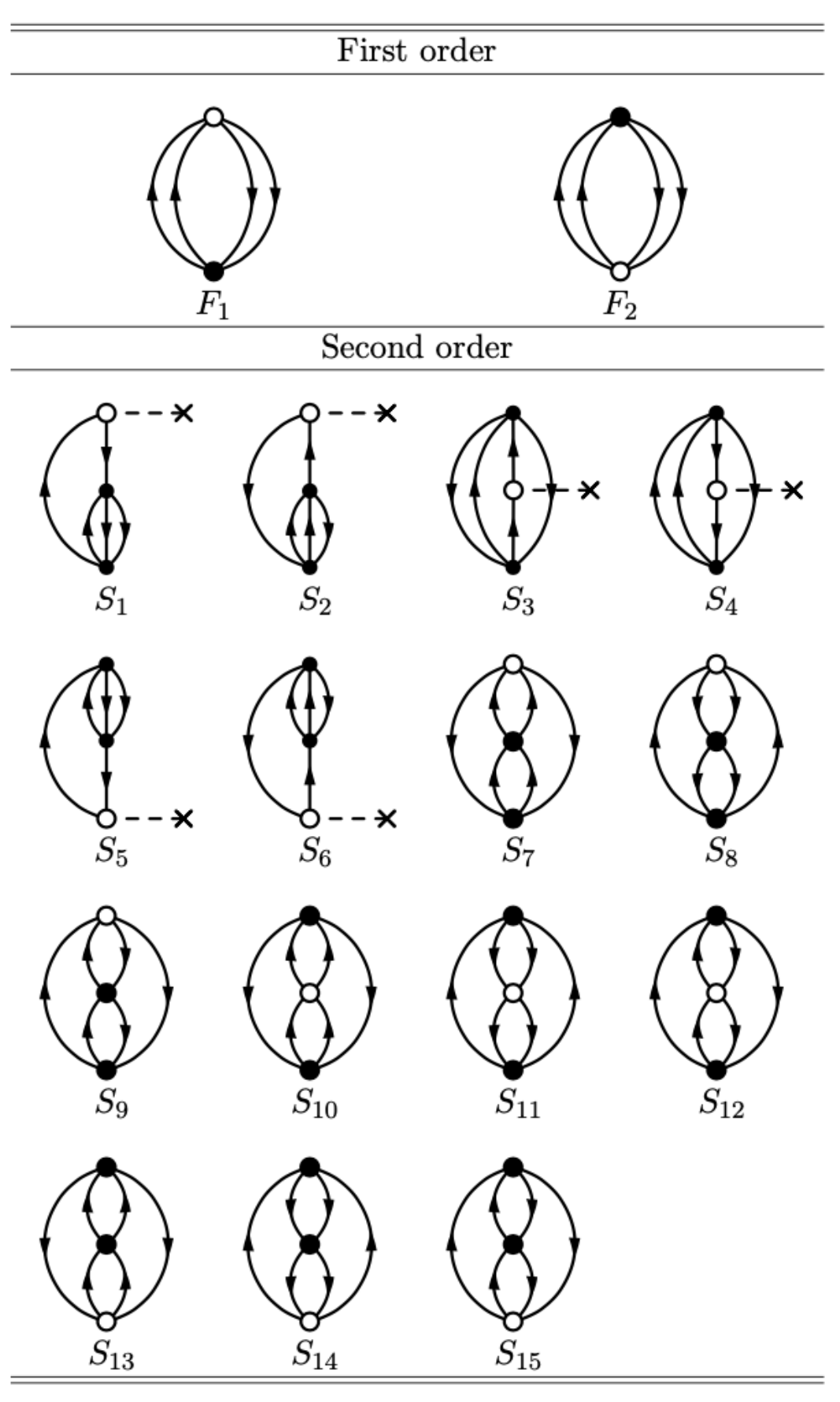}
    \caption{Hugenholtz diagrams for the ground-state expectation value of a scalar operator up to the second order. The solid and open circles indicate Hamiltonian and scalar operators, respectively. The Hartree-Fock basis is assumed. The diagram rules are same as in Ref.~\cite{Shavitt2009a}.}
    \label{fig:scalar_HFMBPT}
\end{figure}
For the ground-state expectation value of a scalar operator in the second-order HF-MBPT, the diagrams are shown in Fig.~\ref{fig:scalar_HFMBPT}.
The expectation value of the scalar operator $S$ is given as
\begin{equation}
\langle S \rangle \approx \langle {\rm HF}| S | {\rm HF} \rangle + \sum_{i=1}^{2}F_{i} + \sum_{i=1}^{15}S_{i}.
\end{equation}
In actual calculations, we use the efficient $J$-coupled scheme.
Below we provide the explicit expressions corresponding to the diagrams.
Let $S_{pq}$ and $S^{J}_{pqrs}$ be the one- and $J$-coupled two-body matrix elements of the scalar operator, and we use the notation:
\begin{equation}
\begin{aligned}
\bar{X}^{J}_{pqrs} &= \sqrt{(1+\delta_{pq})(1+\delta_{rs})} X^{J}_{pqrs}, \\
\bar{X}^{J, \rm{CC}}_{pqrs} &=
\sum_{J'} [J'] \sixj{j_{p}}{j_{q}}{J'}{j_{r}}{j_{s}}{J}
\bar{X}^{J'}_{psrq}, \\
\epsilon^{ab\ldots}_{ij\ldots} &= (f_{ii} + f_{jj} + \cdots) - (f_{aa} + f_{bb} + \cdots).
\end{aligned}
\end{equation}
Here, $X^{J}_{pqrs}$ is the normalized antisymmetrized two-body matrix element of either Hamiltonian or scalar operator.
In the following, we show the $J$-coupled expressions for the diagrams.
As in the main text, we use the convention that $a,b,c,d$ label particle states and $i,j,k,l$ label hole states.
\begin{equation}
F_{1} = \frac{1}{4} \sum_{abij} \sum_{J}
[J]\frac{\bar{\Gamma}^{J}_{abij} \bar{S}^{J}_{abij}}{\epsilon^{ab}_{ij}}
\end{equation}
\begin{equation}
F_{2} = F_{1}
\end{equation}
\begin{equation}
S_{1} = - \frac{1}{2} \sum_{abijk} \sum_{J}
[J]
\frac{\bar{\Gamma}^{J}_{abij}\bar{\Gamma}^{J}_{kbij}
S_{ak}}{\epsilon^{ab}_{ij} \epsilon^{a}_{k}}
\end{equation}
\begin{equation}
S_{2} = \frac{1}{2} \sum_{abcij} \sum_{J}
[J]
\frac{\bar{\Gamma}^{J}_{abij}\bar{\Gamma}^{J}_{abcj}
S_{ci}}{\epsilon^{ab}_{ij} \epsilon^{c}_{i}}
\end{equation}
\begin{equation}
S_{3} = \frac{1}{2} \sum_{abcij} \sum_{J}
[J]
\frac{\bar{\Gamma}^{J}_{abij}\bar{\Gamma}^{J}_{acij} S_{bc}}{\epsilon^{ab}_{ij} \epsilon^{ac}_{ij}}
\end{equation}
\begin{equation}
S_{4} = -\frac{1}{2} \sum_{abijk} \sum_{J}
[J]
\frac{\bar{\Gamma}^{J}_{abij}\bar{\Gamma}^{J}_{abik} S_{jk}}{\epsilon^{ab}_{ij} \epsilon^{ab}_{ik}}
\end{equation}
\begin{equation}
S_{5} = S_{1}
\end{equation}
\begin{equation}
S_{6} = S_{2}
\end{equation}
\begin{equation}
S_{7} = \frac{1}{8} \sum_{abcdij} \sum_{J}[J]
\frac{\bar{\Gamma}^{J}_{abij} \bar{\Gamma}^{J}_{abcd}
\bar{S}^{J}_{cdij}}{\epsilon^{ab}_{ij}\epsilon^{cd}_{ij}}
\end{equation}
\begin{equation}
S_{8} = \frac{1}{8} \sum_{abijkl} \sum_{J}[J]
\frac{\bar{\Gamma}^{J}_{abij} \bar{\Gamma}^{J}_{jikl}
\bar{S}^{J}_{abkl}}{\epsilon^{ab}_{ij}\epsilon^{ab}_{kl}}
\end{equation}
\begin{equation}
S_{9} = -\sum_{abcijk}
\sum_{J} [J]
\frac{\bar{\Gamma}^{J, {\rm CC}}_{ajib} \bar{\Gamma}^{J, {\rm CC}}_{ibkc}
\bar{S}^{J, {\rm CC}}_{kcaj}}
{\epsilon^{ab}_{ij}\epsilon^{ac}_{jk}}
\end{equation}
\begin{equation}
S_{10} = \frac{1}{8} \sum_{abcdij} \sum_{J}[J]
\frac{\bar{\Gamma}^{J}_{abij} \bar{S}^{J}_{abcd}
\bar{\Gamma}^{J}_{cdij}}{\epsilon^{ab}_{ij}\epsilon^{cd}_{ij}}
\end{equation}
\begin{equation}
S_{11} = \frac{1}{8} \sum_{abijkl} \sum_{J}[J]
\frac{\bar{\Gamma}^{J}_{abij} \bar{S}^{J}_{jikl}
\bar{\Gamma}^{J}_{abkl}}{\epsilon^{ab}_{ij}\epsilon^{ab}_{kl}}
\end{equation}
\begin{equation}
S_{12} = -\sum_{abcijk}
\sum_{J}[J]
\frac{\bar{\Gamma}^{J, {\rm CC}}_{ajib} \bar{S}^{J, {\rm CC}}_{ibkc}
\bar{\Gamma}^{J, {\rm CC}}_{kcaj}}
{\epsilon^{ab}_{ij}\epsilon^{ac}_{jk}}
\end{equation}
\begin{equation}
S_{13} = S_{7}
\end{equation}
\begin{equation}
S_{14} = S_{8}
\end{equation}
\begin{equation}
S_{15} = S_{9}
\end{equation}

\end{document}